\documentclass[journal]{IEEEtran}
\usepackage{balance}
\usepackage[ruled]{algorithm2e}
\usepackage{amssymb}
\usepackage{cite}
\usepackage{color}
\usepackage{multirow}
\usepackage{graphicx,times,amsmath}
\usepackage{subfigure}

\usepackage{amssymb}
\usepackage{url}
\ifCLASSINFOpdf

\else

\fi

\begin{document}
\title{Sensor-Augmented Neural Adaptive Bitrate Video Streaming on UAVs} 
	
\author{Xuedou~Xiao,~Wei~Wang,~\IEEEmembership{Member,~IEEE,}~Taobin~Chen,~Yang~Cao,~\IEEEmembership{Member,~IEEE,}~Tao~Jiang,~\IEEEmembership{Fellow,~IEEE,}~Qian~Zhang,~\IEEEmembership{Fellow,~IEEE}
		
\thanks{X. Xiao, W. Wang, T. Chen, Y. Cao and T. Jiang are with the School of Electronic Information and Communications, Huazhong University of Science and Technology, Wuhan, China (e-mail: xuedouxiao@hust.edu.cn; weiwangw@hust.edu.cn; chentaobin@hust.edu.cn; ycao@hust.edu.cn; taojiang@hust.edu.cn).} 
\thanks{Q. Zhang is with the Department of Computer Science and Engineering, Hong Kong University of Science and Technology, Clear Water Bay, Hong Kong (e-mail:qianzh@cse.ust.hk).}	}
\maketitle
	
\begin{abstract} 
Recent advances in unmanned aerial vehicle (UAV) technology have revolutionized a broad class of civil and military applications. However, the designs of wireless technologies that enable real-time streaming of high-definition video between UAVs and ground clients present a conundrum. Most existing adaptive bitrate (ABR) algorithms are not optimized for the air-to-ground links, which usually fluctuate dramatically due to the dynamic flight states of the UAV. In this paper, we present SA-ABR, a new sensor-augmented system that generates ABR video streaming algorithms with the assistance of various kinds of inherent sensor data that are used to pilot UAVs. By incorporating the inherent sensor data with network observations, SA-ABR trains a deep reinforcement learning (DRL) model to extract salient features from the flight state information and automatically learn an ABR algorithm to adapt to the varying UAV channel capacity through the training process. SA-ABR does not rely on any assumptions or models about UAV's flight states or the environment, but instead, it makes decisions by exploiting temporal properties of past throughput through the long short-term memory (LSTM) to adapt itself to a wide range of highly dynamic environments. We have implemented SA-ABR in a commercial UAV and evaluated it in the wild. We compare SA-ABR with a variety of existing state-of-the-art ABR algorithms, and the results show that our system outperforms the best known existing ABR algorithm by 21.4\% in terms of the average quality of experience (QoE) reward.

\end{abstract}


\vspace{0.3cm}
\section{Introduction}\label{sec:intro} 

Inexpensive commercially available unmanned aerial vehicles (UAVs) are rising rapidly, making drones a popular host of a wide class of applications, including environment monitoring~\cite{monitor}, precision agriculture~\cite{agri}, photography~\cite{Fleureau2016POSTERGD}, policing~\cite{VzrdQf}, firefighting~\cite{WwFgWk}, and package delivery~\cite{timmurphy}. An essential functionality enabling these applications is to record high-definition videos of superior quality and seamlessly share them with ground base stations or clients for manual inspection and further analysis.



Despite the excitement, today's UAVs are struggling to deliver high-quality video in real time to ground receivers. Today's commercial UAVs adopt fixed-bitrate video streaming strategies which may result in severe rebuffering under poor channel conditions~\cite{h520}. In addition, many studies have adopted various kinds of adaptive bitrate (ABR) algorithms~\cite{Stockhammer2011Dynamic,Jiang2012Improving,Sun2016CS2P,Huang2014A,Spiteri2016BOLA,Zhi2014Probe,Yin2015A,KimXMAS,zhou2016mdash,Akhtar:2018:OAV:3230543.3230558,xie2017dynamic,xu2018qoe,Claeys2013Design,Chiariotti2016Online,Claeys2014Design,Hooft2015A,Mao2017Neural,Huang:2018:QVQ:3240508.3240545,Yeo:2018:NAC:3291168.3291216,jiang2018chameleon,sengupta2018hotdash,guo2019buffer,kan2019deep}, including learning methods~\cite{Claeys2013Design,Chiariotti2016Online,Claeys2014Design,Hooft2015A,Mao2017Neural,Huang:2018:QVQ:3240508.3240545,Yeo:2018:NAC:3291168.3291216,jiang2018chameleon,sengupta2018hotdash,guo2019buffer,kan2019deep}, under various network conditions on the ground. These algorithms make ABR decisions based on network observations and video playback states. However, it is challenging for them to fit well in 
UAV communications, as the channel capacity of air-to-ground links fluctuate dramatically and the primary causes come from factors including varying environments and dynamic motion states, such as flight velocities, intense vibrations and distances from the ground clients. These factors result in unique patterns in the channel capacity variances, which can hardly be learned from models built for ground-to-ground links. Consequently, the majority of ABR algorithms either fail to transmit higher-quality video streaming~\cite{Jiang2012Improving} or exceed the channel capacity~\cite{Huang2014A} due to unexpected situations. 
To break this stalemate, dedicated models tailored for air-to-ground links are required to cope with such unique variance patterns.

Instead of solely relying on the network observations and video playback states which are not sufficient enough to adapt to the highly dynamic air-to-ground links, we argue to incorporate more inherent sensor data that can reflect UAV flight states in the ABR algorithm. Through field tests and measurements, we observe that the UAV's flight-state-related sensor data can provide hints about channel variance patterns, which can guide the design of ABR algorithms. In addition, the temporal patterns in the past throughput variances can be further extracted and exploited to obtain valuable information about current channel conditions. Thus, we believe it is essential to incorporate the flight-state-related sensor data as side information to design video streaming strategies for UAVs.

\begin{figure}[t]
	\centering
	\includegraphics[width=0.48\textwidth]{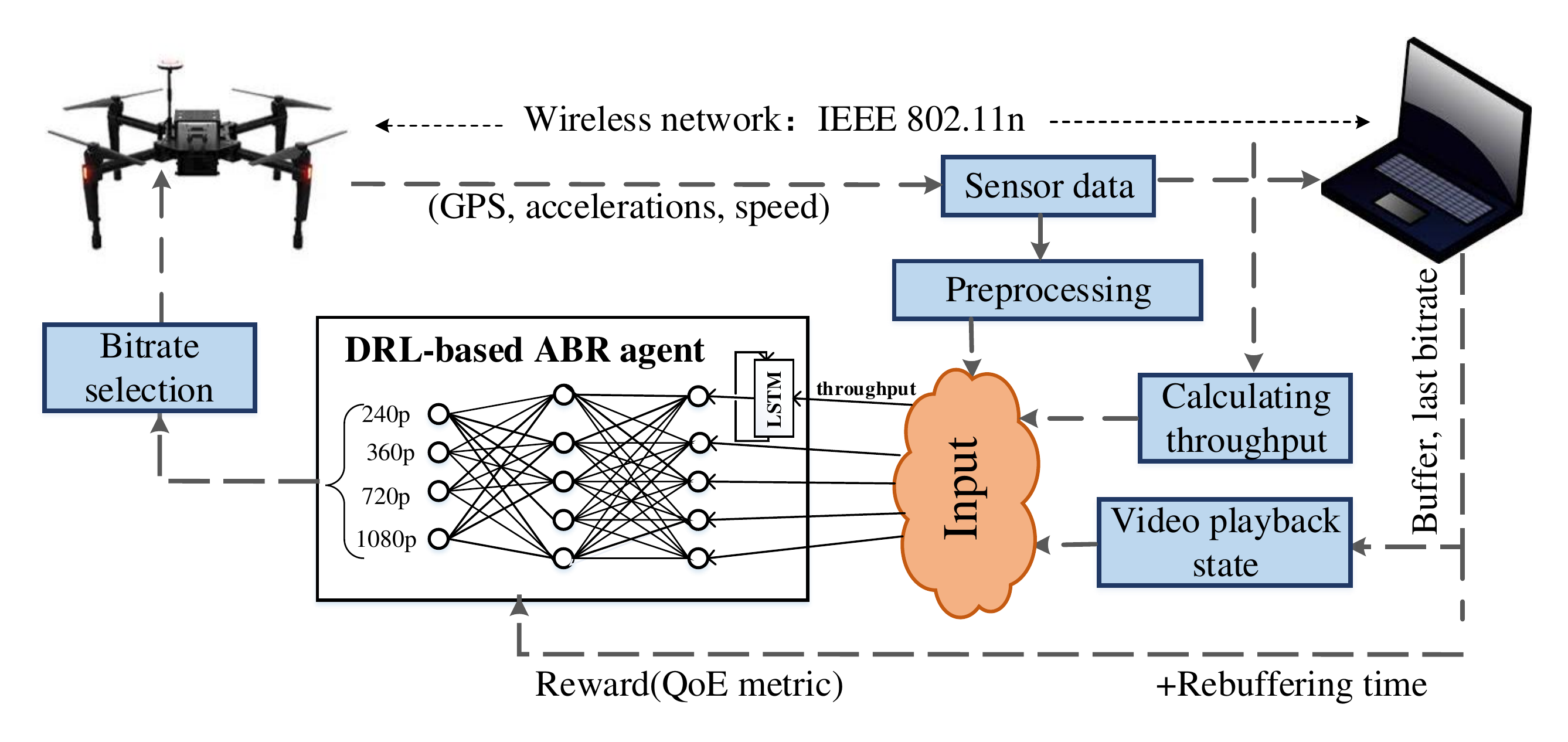}
	\caption{An overview of SA-ABR architecture.}\label{design}
\end{figure}

In this work, we present \textit{SA-ABR}, a new sensor-augmented system that generates ABR video streaming algorithms based on deep reinforcement learning (DRL)~\cite{goodfellow2016deep,sutton2018reinforcement}, which aims at obtaining optimal bitrate selection strategies under varying UAV channel conditions. The goal of SA-ABR is to maximize the video quality of experience (QoE)~\cite{ge2017towards,lu2018exploiting} for viewers
through the training process. As illustrated in Fig.~\ref{design}, apart from the state information of past throughput experience and video playback, SA-ABR also feeds the inherent sensor data, including GPS coordinates, acceleration and velocity, which indicates the flight state of the UAV, into the neural network. The sensor data is updated in real time at the beginning of each video chunk. Additionally, the average throughput of past video chunks is calculated by recording the number of packets. 
The focus of our model is on how to capture the salient features from the time-series throughput sequences and make full use of the sensor data to better estimate the future throughput trend. After the training process, our model can automatically adapt to the throughput dynamics and make optimal bitrate decisions for the next video chunks. 

A key challenge in our design is how to make full use of the sensor data to extract features that are indicative of throughput dynamics. Although there are some relations between the sensor value and the throughput, it is prohibitively complex to describe these relations in analytical expressions or clear rules. For example, the acceleration pattern is generally very misleading due to the existence of the UAVs' own vibrations, making it intractable for the DRL model to distinguish whether the acceleration data comes from the change of flight states or vibrations. To overcome this hurdle, SA-ABR applies quantization-based preprocessing to sensor data before directly feeding it to the neural network. This not only ensures the full use of the sensor data that provide hints about channel conditions, but also eliminates irrelevant noise and disturbance. Another obstacle stems from how to effectively analyze the temporal characteristics of throughput when making predictions about future throughput. 
We incorporate the long short-term memory network (LSTM)~\cite{Memory2010Long} into the DRL model, which exploits the unique memory function of the LSTM to better capture the temporal properties and improve the accuracy of throughput forecasts.






We implement SA-ABR on a DJI Matrice 100 and compare the performance of our system with state-of-the-art ABR algorithms in the wild. The results show that SA-ABR achieves up to a 21.4$ \% $ gain in the average QoE reward over the best known existing ABR algorithm~\cite{Mao2017Neural}.


The main contributions of this paper are summarized as follows.
\begin{itemize}
	\item We conduct a comprehensive measurement study to explore the impact of UAV's motions on throughput, which provides hints to optimize the video streaming strategies from the flight-state perspective. 
	\item We propose a new DRL-based ABR architecture with the assistance of the sensor data. The model exploits the inherent sensor data that is used to pilot UAVs to better adapt to the highly dynamic air-to-ground links.
	\item We implement our design on a commercial UAV and conduct a series of experiments in the wild to validate our system. The results demonstrate the feasibility of adapting to the air-to-ground channel dynamics, resulting in a 21.4$ \% $ increase in the average QoE reward compared to the best known existing ABR algorithm.
\end{itemize}

The remainder of this paper is structured as follows. We begin in Section~\ref{sec:mobility} to explore the impact of the UAV's flight states. Section~\ref{sec:training} describes the system design of our sensor-augmented DRL algorithm.
Section~\ref{sec:implementation} describes the system implementation and Section~\ref{sec:evaluation} evaluates the performance of SA-ABR. Sections~\ref{sec:related work} and \ref{sec:conclusion} present the related work and conclusion.

\vspace{0.2cm}
\section{Exploiting Fight State Awareness}\label{sec:mobility}


In this section, we start by giving a brief introduction of the flight-state-related sensor data available on UAVs. Then, we conduct experiments in controlled flight states to reveal the specific relationship between throughput and the sensor data. Finally, we proceed to take a deeper inspection of the complex relationship and summarize counter-intuitive observations and irregular phenomena, which motivates our DRL-based design.

\subsection{Flight States of the UAV}

UAVs such as multi-rotor drones need to identify their flight states at all times, including 3D position, 3D orientation, and their derivatives. Therein, the positions and velocities are important data for the UAV to determine whether it is hovering or moving. As UAVs normally adjust the postures to generate thrusts in certain directions, they also need to measure their own current 3D orientations in real time. Together with corresponding accelerations and angular velocities, there are a total of fifteen state quantities that are required to identify the flight states. 
With various sensor technologies, including GPS, inertial measurement unit (IMU), barometer and geomagnetic compass, equipped on UAVs, we can obtain all fifteen state quantities that comprehensively reflect the flight conditions. To explore how the flight states affect the air-to-ground channel conditions, we collect the sensor data, including GPS coordinates, velocities, accelerations and further analyze the relationship between sensor data and throughput. 

\subsection{Impact of Flight States}
As the capability of throughput forecast plays an essential role in the ABR video streaming algorithm to make bitrate selection meeting viewers' QoE requirements, we start with a series of tests to analyze the underlying relationship between the sensor data and the throughput in controlled flight states. In these tests, the transmitter is a DJI Matrice 100 UAV and sends the data file through the IEEE 802.11n protocol. The ground receiver uses the commercial WiFi network card with a USB interface embedded on the laptop.

The sites of measurements are selected to validate the generalization of our observations, including a playground, a plaza and a pool on campus, each with its own channel characteristics that cover a majority of different scenarios. The air-to-ground propagation on the playground can be described as an ordinary two-ray model~\cite{Khuwaja2018ASO}. Different from the unobstructed playground, the plaza is surrounded by buildings and trees. Under such circumstances, the propagation is affected by the shadowing losses and the multipath diversities. Moreover, the surface reflectivity and roughness of the pool are different from the ground, resulting in varying parameters in propagation models. For each flight state, we collect data lasting more than 150~s in each place.

\textbf{Impact of distance.} We perform a set of tests to analyze the impact of the distance between the ground client and the UAV on throughput. This set of experiments includes two groups of measurements according to the UAV's flight states, i.e., the hovering and moving states. In the hovering experiment, we collect throughput and sensor data when the UAV hovers at a height of 10~m and the link distances vary from 10~m to 60~m. In the moving experiment, we consider a higher altitude of 20~m for safety reasons, with various distances ranging from 20~m to 60~m. In addition, the UAV is controlled to fly at different distances at a constant velocity of 8~m/s.

\begin{figure}
	\centering
	\subfigure[Throughput vs. distance in hovering state.]{
		\includegraphics[width=0.48\linewidth]{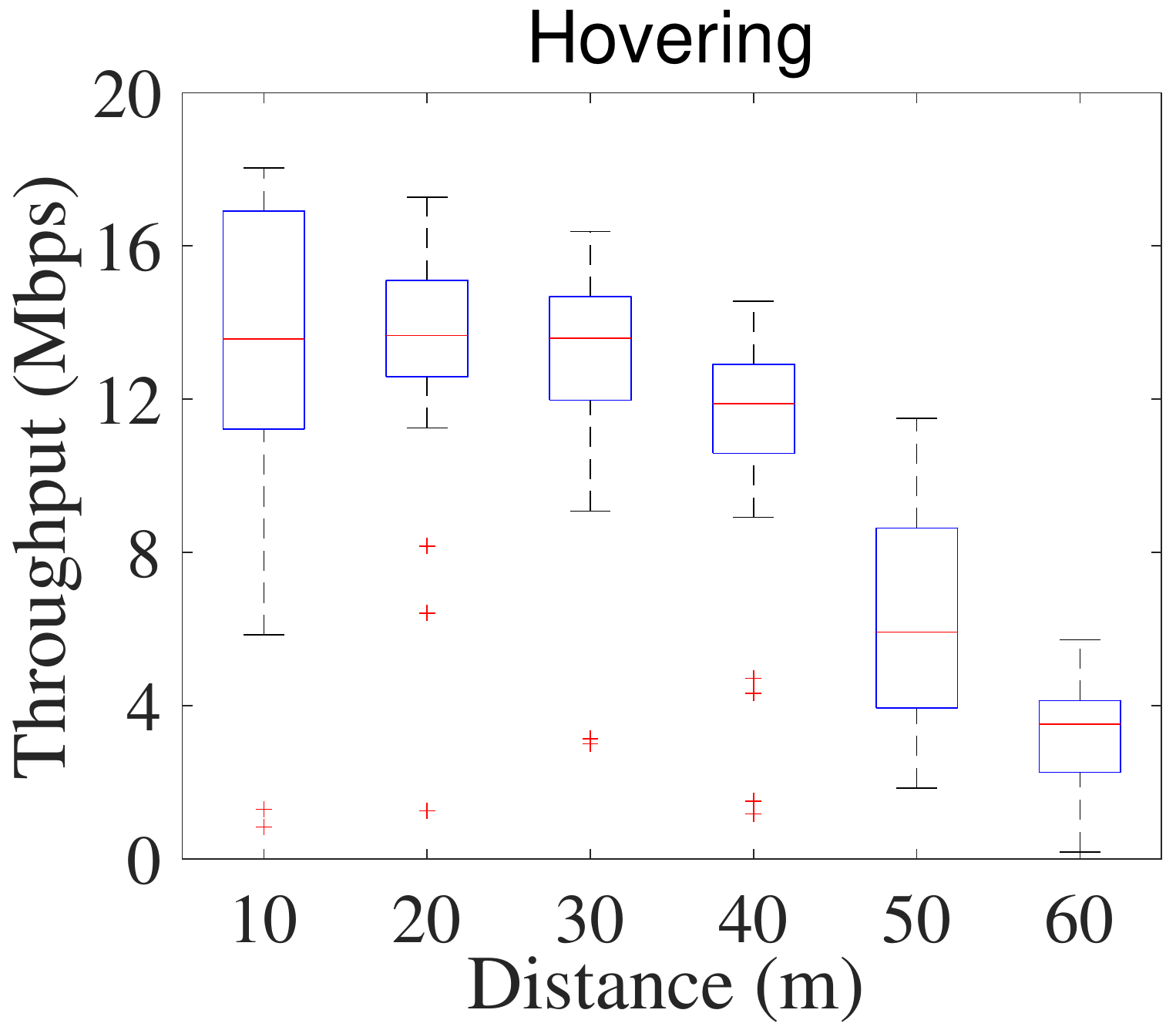}\label{2a}}
	\hfill
	\subfigure[Throughput vs. distance at an average speed of 8~m/s.]{
		\includegraphics[width=0.48\linewidth]{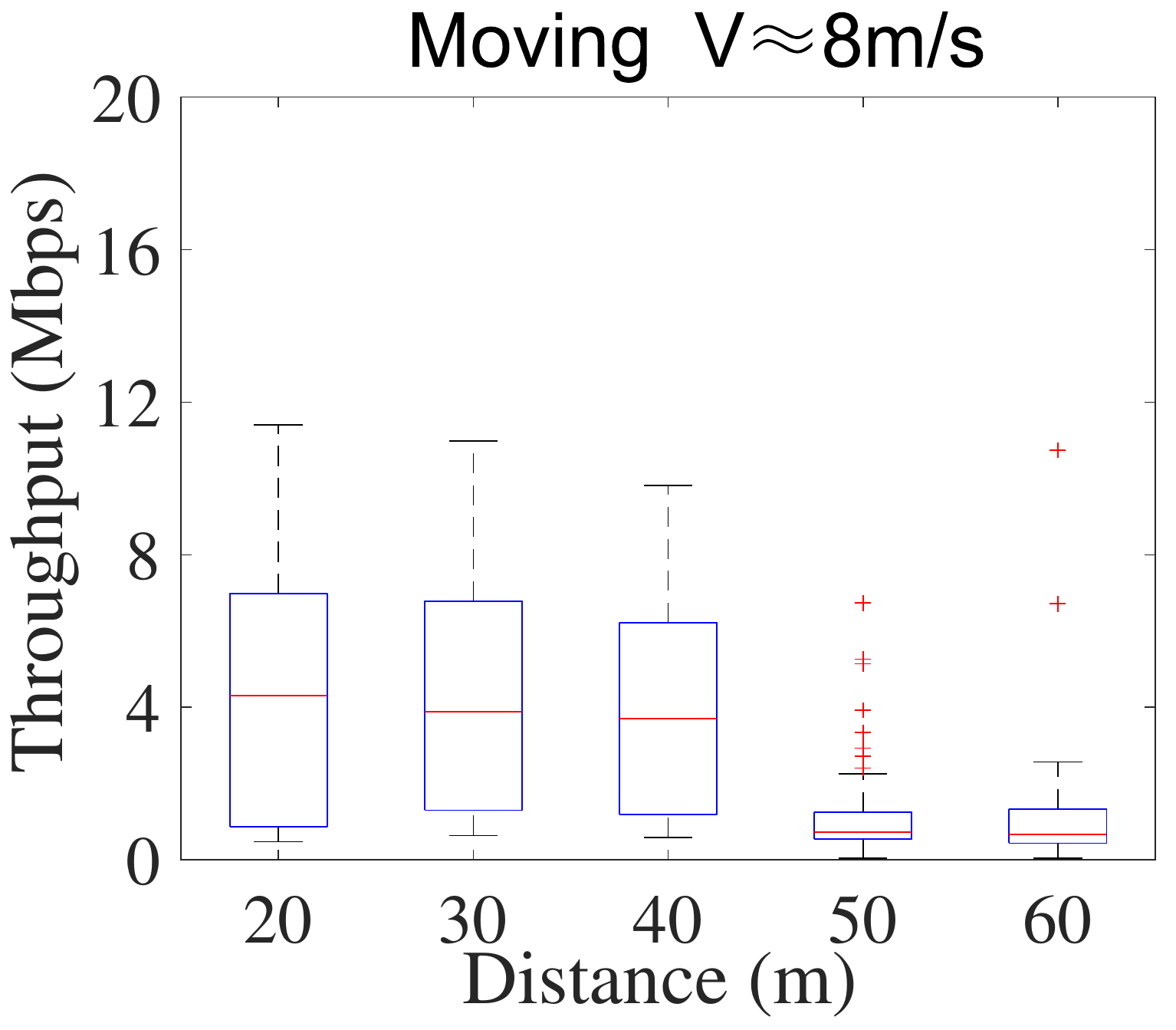}\label{2b}}
	\\
	\caption{Throughput vs. distance in controlled flight states.}
	\label{2} 
\end{figure}

The experimental results are presented in Fig.~\ref{2}. 
From the mediums, the quartiles and the ends of each boxplot, we can conclude that the throughput decreases with distances no matter whether the UAV is moving or hovering. This rule provides a basis for SA-ABR to incorporate the distance value that affects the throughput, into the algorithm.


\begin{figure}
	\centering
	\subfigure[Throughput vs. velocity at an average distance of 20~m.]{
		\includegraphics[width=0.48\linewidth]{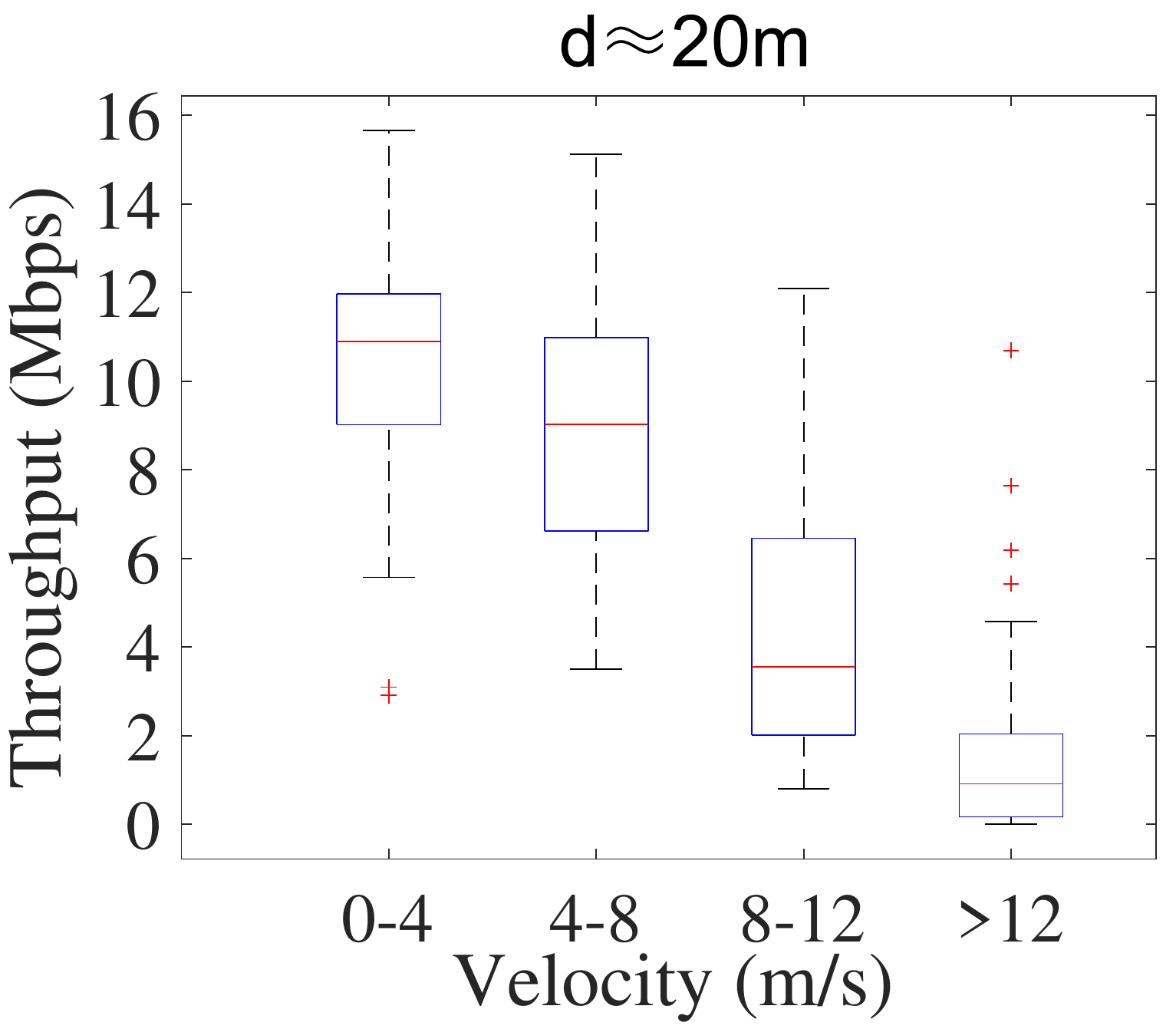}\label{3a}}
	\hfill
	\subfigure[Throughput vs. velocity at an average distance of 50~m.]{
		\includegraphics[width=0.48\linewidth]{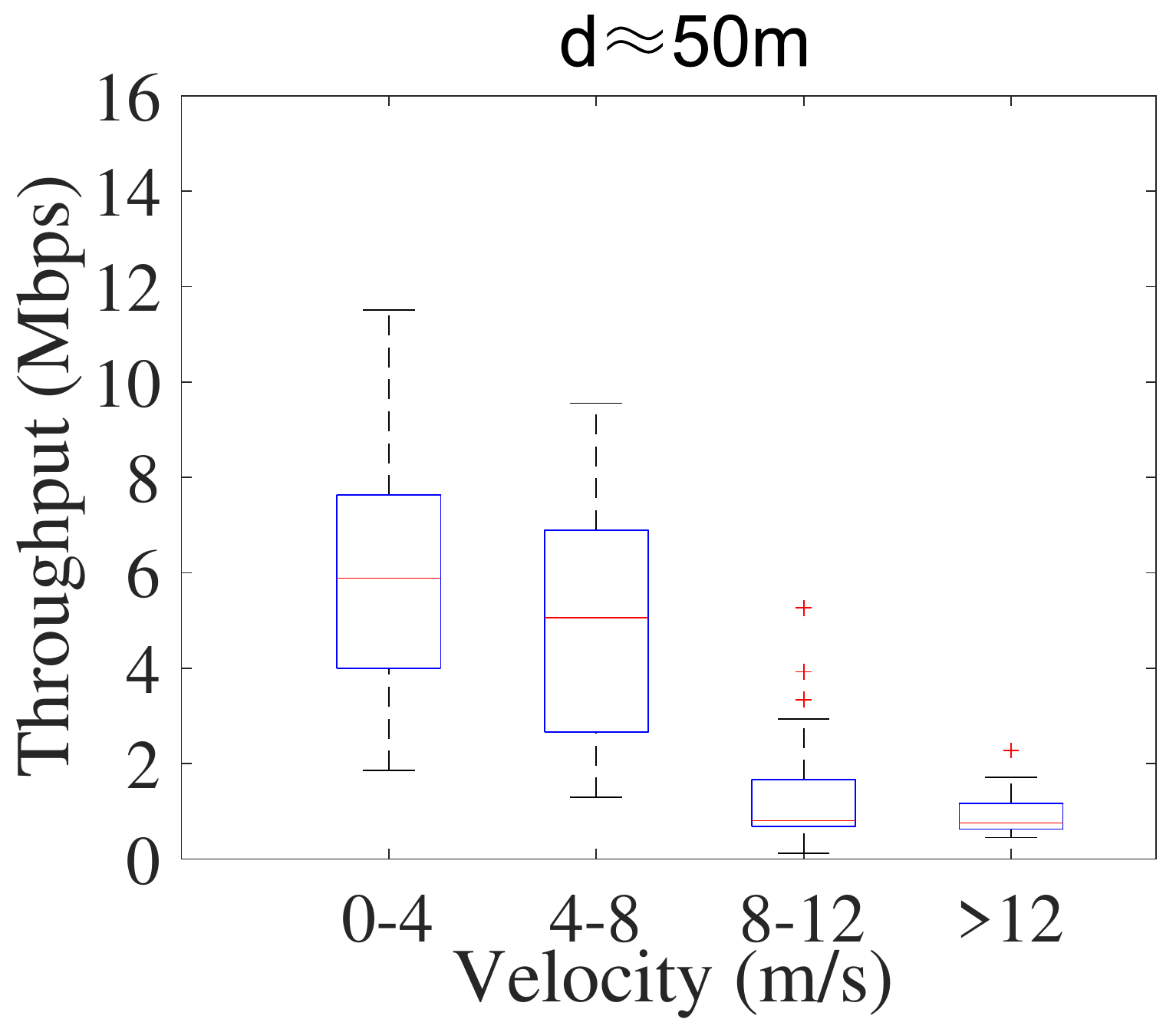}\label{3b}}
	\\
	\caption{Throughput vs. velocity in controlled flight states.}
	\label{3} 
\end{figure}

\textbf{Impact of velocity.} The experiments with different velocities are also conducted by dividing the testing process into two sets: one is at a distance of around 20~m and the other is around 50~m. We control the UAV to fly around the ground client at stable distances while the velocity sweeps from 0 to 16~m/s.


The throughputs achieved at different velocity ranges (including 0-4~m/s, 4-8~m/s, 8-12~m/s, $\textgreater$12~m/s) are shown in Fig.~\ref{3}. The throughput diminishes quickly when the velocity range increases, no matter whether the link distance is around 20~m or 50~m. Based on the impact of velocity on throughput, we can exploit the velocity data in our algorithm design to better forecast throughput.



\textbf{Impact of acceleration.} To obtain a better bitrate adaptation model, we further perform tests and investigate the impact of acceleration. The acceleration data includes three dimensions: $a_x$, $a_y$, $a_z$. During the flight, the UAV continuously accelerates and decelerates while the distance from the ground client remains stable.

\begin{figure}[t]
	\centering
	\includegraphics[width=0.49\textwidth]{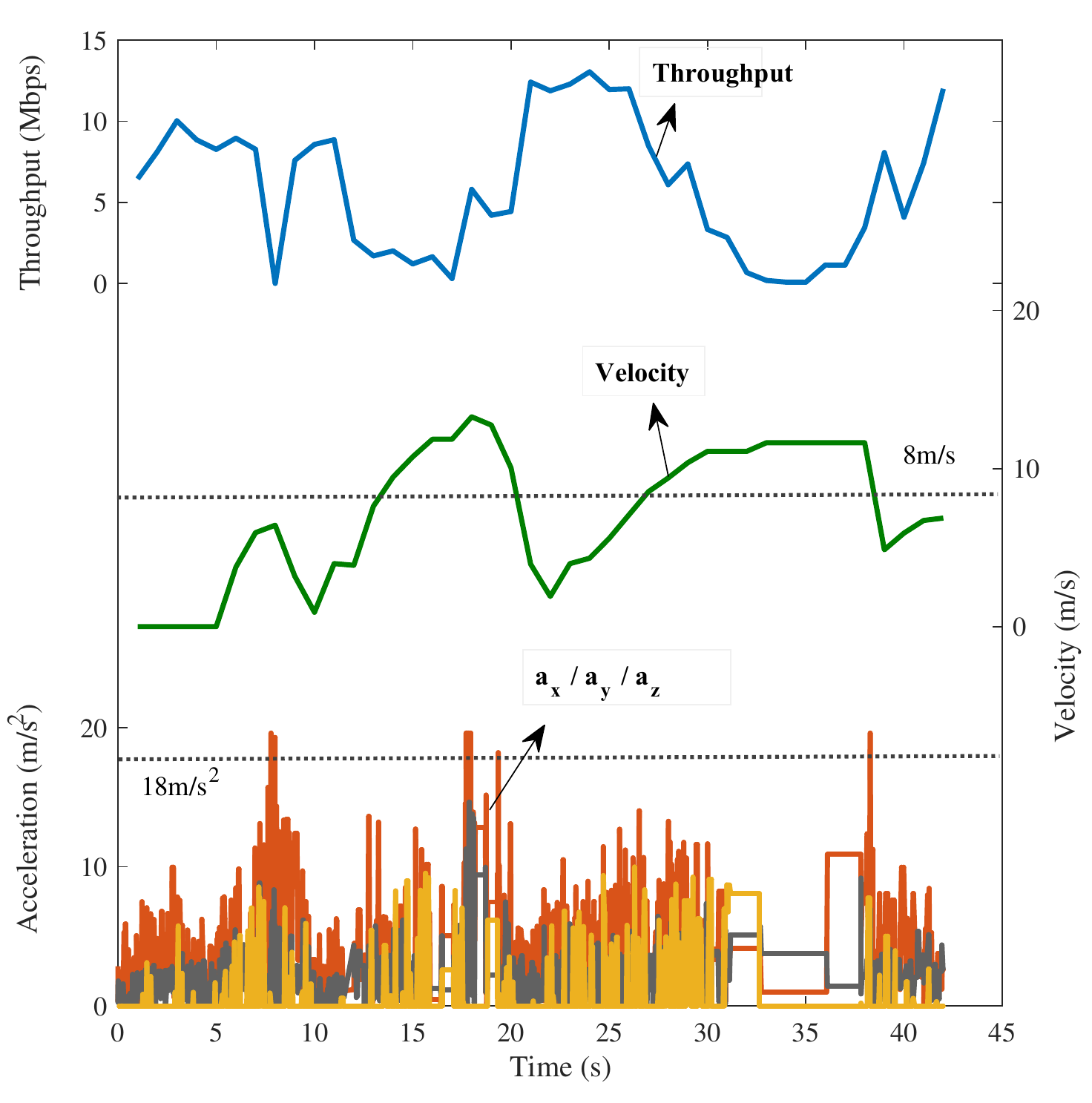}
	\caption{Profiling data patterns including three-dimensional accelerations, velocities and throughput.}\label{acce_zong}
\end{figure}

The results are demonstrated in Fig.~\ref{acce_zong}. 
For the acceleration data, the low value is dominated by the UAV vibrations, while the high value can indicate the changes in flight states. Thus, we can pick out the high values of acceleration data to analyze their impact on throughput. 
Note that the variances in acceleration and velocity are synchronous to some extent, which may compromise the impact of acceleration data on throughput prediction.
Nevertheless, when the acceleration increases rapidly at 8s, the velocity value has not reached large enough to determine if the channel quality becomes worse. Therefore, the utilization of acceleration data can make up for this weakness in our algorithm and serve as a supplement to velocity data to perform better throughput forecast.


\textbf{Summary.} The general relationships between the sensor data and the throughput in the controlled flight states can be summarized as follows: (1) the throughput monotonically decreases as the distance and the speed increase, and (2) the acceleration fluctuations are able to reflect the throughput changes, i.e., the larger the variance in acceleration data, the more likely the throughput will be affected. These rules exist due to complex latent factors including the path loss, the multipath effects, the modulation coding scheme (MCS) conversions and the fast changes in environments caused by the high-speed movements of the UAV. Thus, the flight-state-related sensor data is indicative of channel state information, which provides hints for SA-ABR to enhance the forecast capability and achieve the best QoE from the flight-state perspective.

\subsection{Deeper Inspection}

The above experiments are conducted with preset or controlled flight states. In this part, we proceed to conduct experiments where the UAV is allowed to fly on random paths at arbitrary speeds in various environments (see Fig.~\ref{uav}), to take a deeper investigation of the relationship between throughput and sensor data. The analyses include three aspects: throughput vs. distance, velocity, and acceleration, respectively. 

\begin{figure}
	\centering
	\subfigure[Throughput vs. distance.]{
		\includegraphics[width=0.479\linewidth]{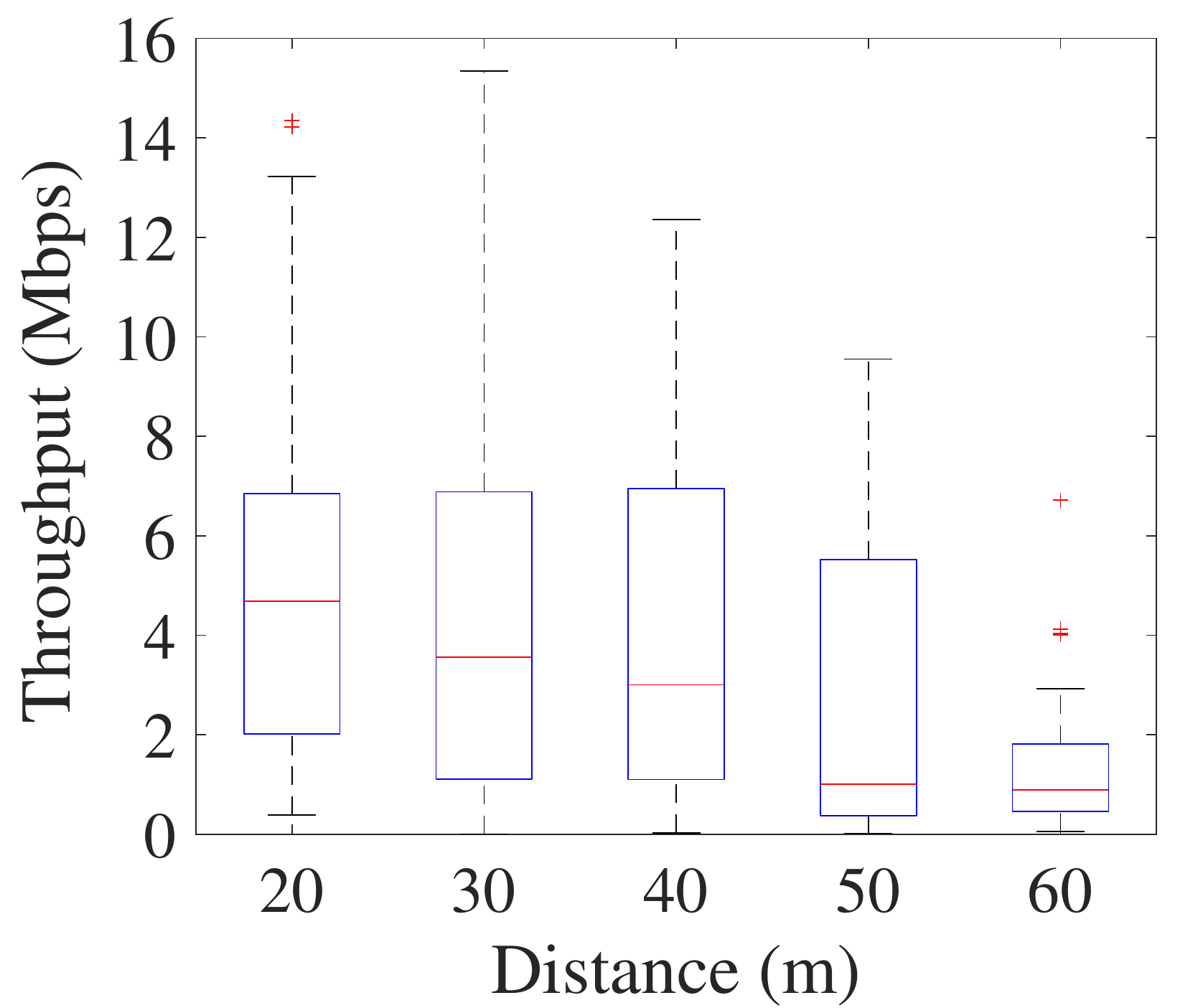}\label{4a}}
	\hfill
	\subfigure[Throughput vs. distance in the from of CDF.]{
		\includegraphics[width=0.472\linewidth]{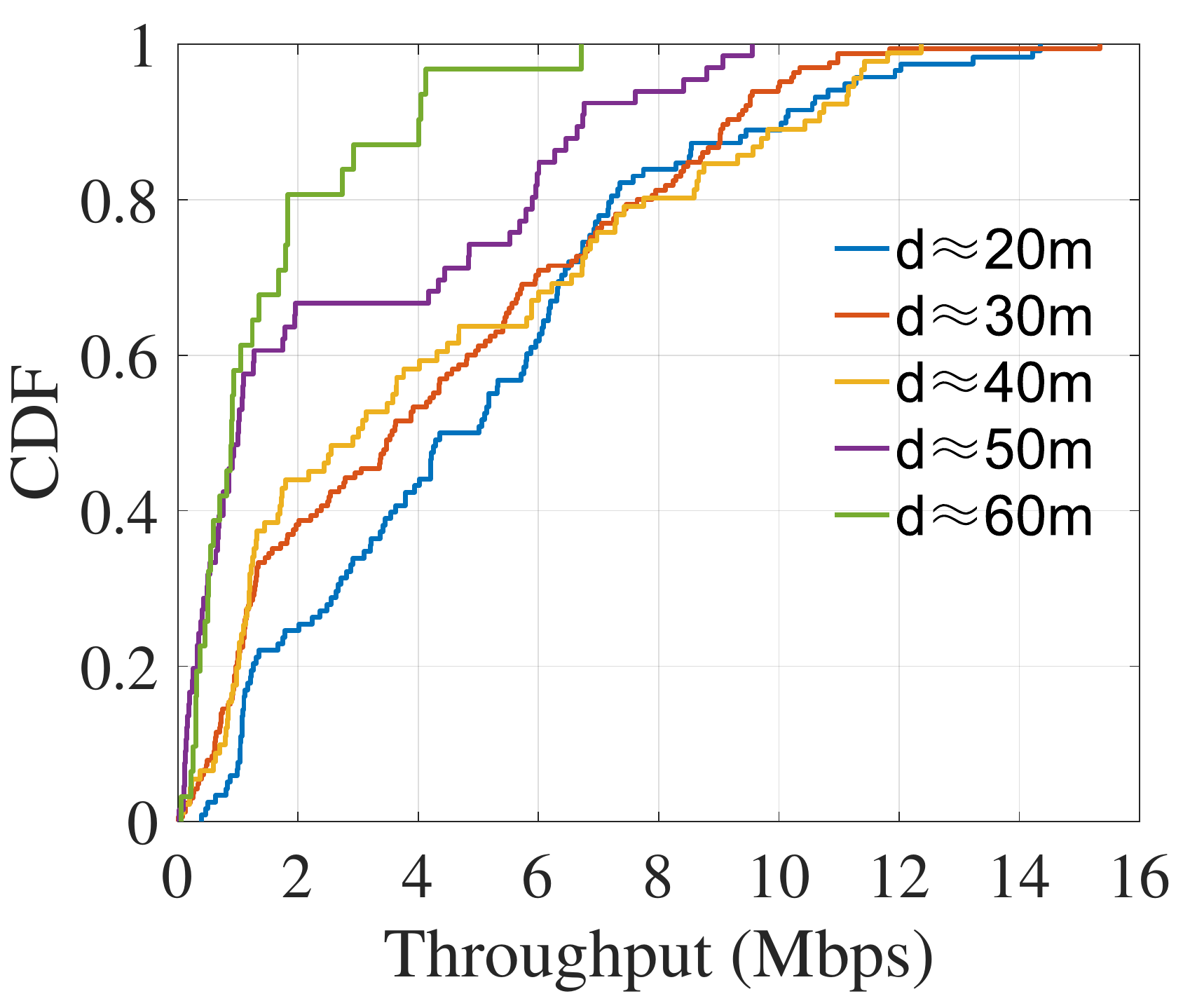}\label{4b}}
	\hfill
	\subfigure[Throughput vs. velocity.]{
		\includegraphics[width=0.465\linewidth]{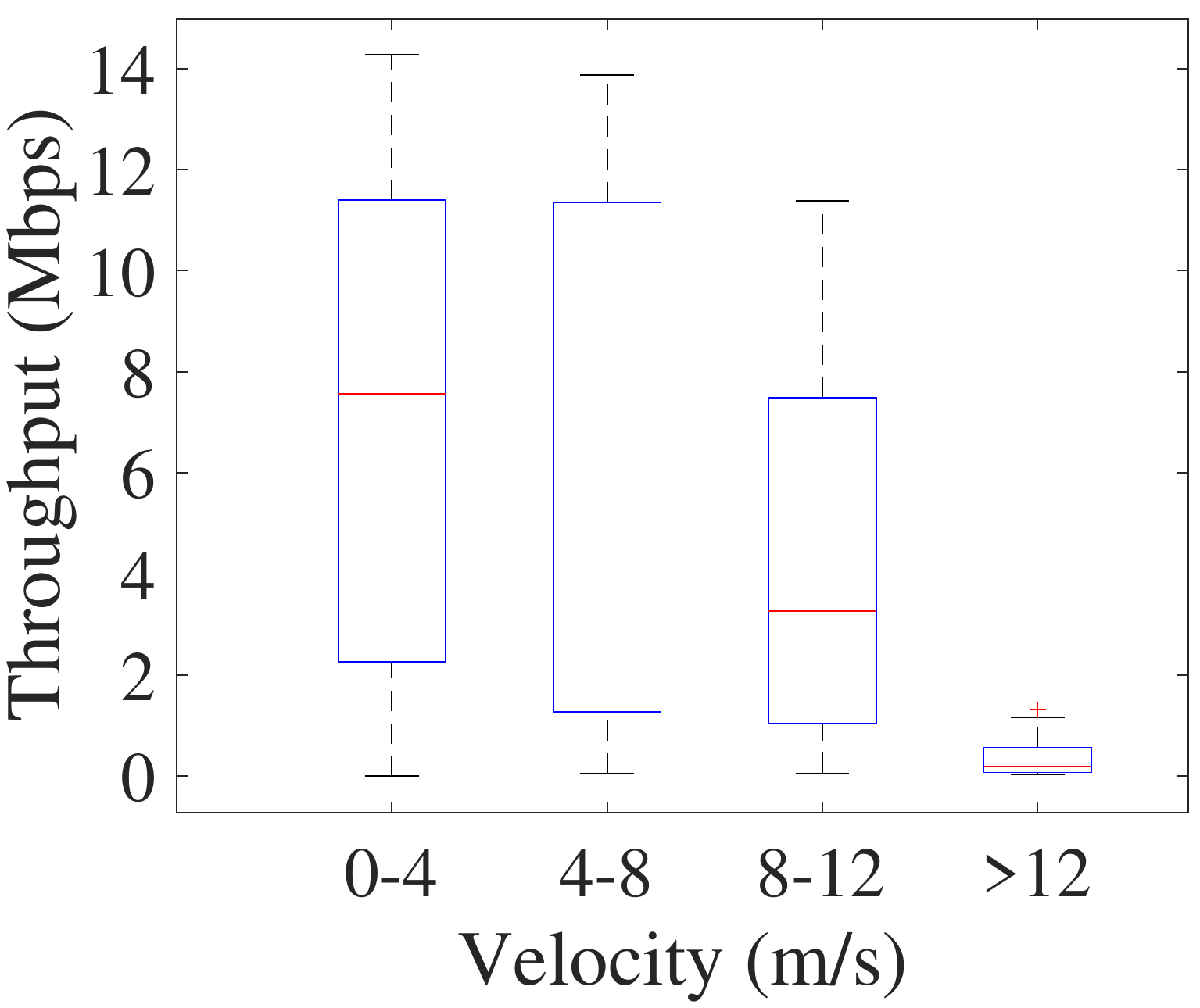}\label{4c}}
	\hfill
	\subfigure[Throughput vs. velocity in the from of CDF.]{
		\includegraphics[width=0.4663\linewidth]{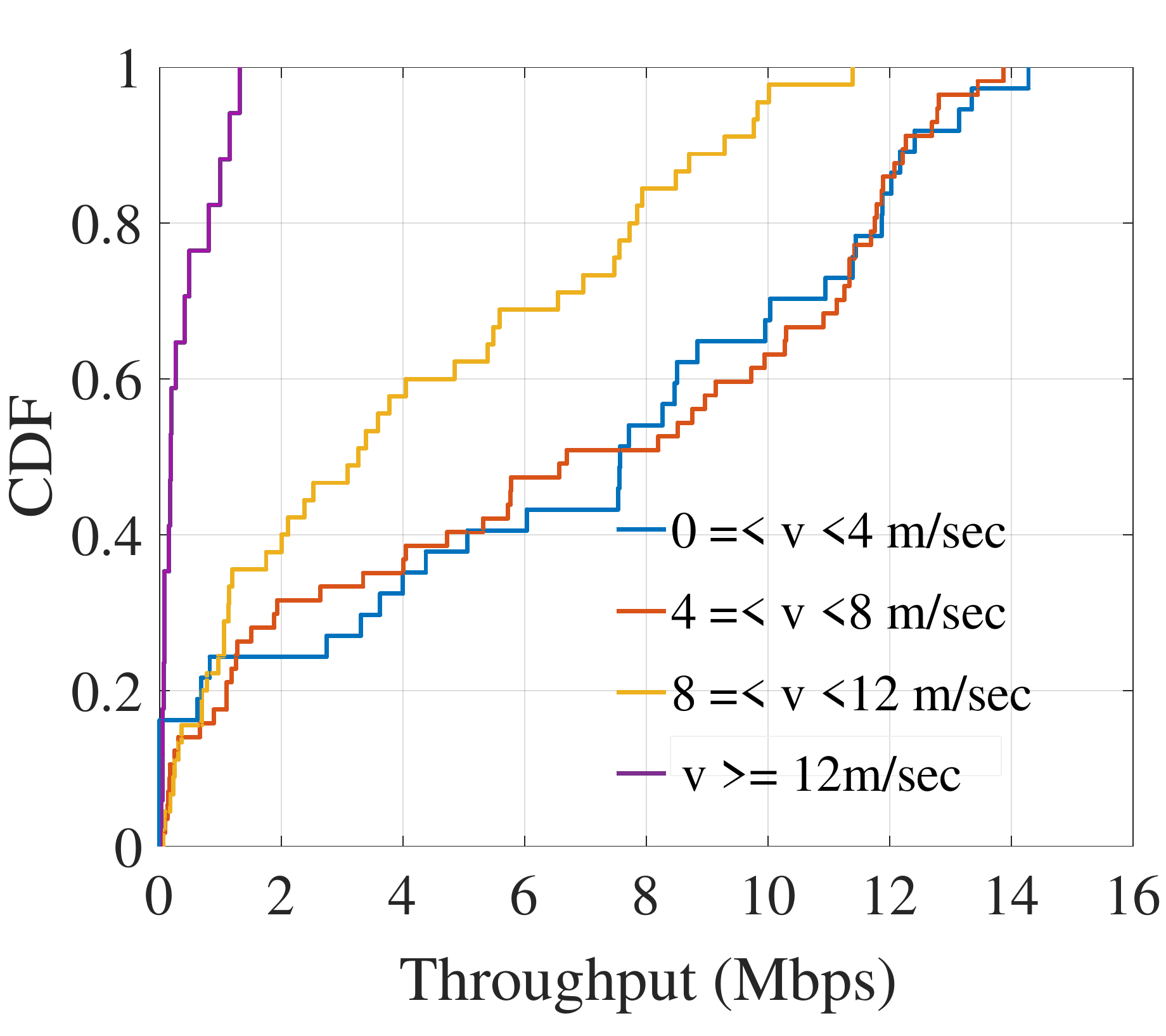}\label{4d}}
	\\
	\caption{Throughput vs. distance and velocity in uncontrolled flight states.}
	\vspace{0.1cm}\label{4} 
\end{figure}

The detailed results of throughput vs. distance and throughput vs. velocity are visualized in Fig.~\ref{4}. We first observe the general rule in Fig.~\ref{4a}, i.e., the throughput decreases with velocities. However, the throughput value at lower distances covers almost the entire range (0-16~Mbps), as indicated in the bottom and top edges of the boxplots. The reason  for this phenomenon may be the uncertainties caused by the rapidly changing velocities of the UAV in the wild. 

In addition, the cumulative distribution function (CDF) in Fig.~\ref{4b} further demonstrates that for lower distances, the throughput is more likely to be distributed in the higher numerical area and less in the lower area, while the throughput at high distances follows an opposite pattern. This phenomenon transforms the regular impact characteristics of distances into a probabilistic problem with uncertainty due to the changing velocities, which is more complex than what we observed in previous experiments. Similar phenomena are also observed in the experiments regarding velocity (see Fig.~\ref{4c} and Fig.~\ref{4d}).


After deeper inspection, some counter-intuitive observations can also be noted in the acceleration data. 
For example, during the deceleration phase (see Fig.~\ref{acce_zong} at around 38~s), the velocity curve drops and the throughput shows an upward trend. The acceleration data, however, shows a sharp fluctuation due to the sudden change in velocity. These irregular phenomena occasionally happen when the impact of acceleration data on throughput contends with other influencing factors, such as distance and velocity, which might result in a misjudgment for the throughput predictor.


\textbf{Summary.} Based on these irregular phenomena and counter-intuitive observations, we observe that the monotonicity in relationships between sensor data and throughput is not as clear as the experiments in controlled flight states. For example, the UAVs at low velocities do not necessarily lead to high-quality wireless channel due to the dominant influence of path loss at long link distance. In other words, various sensor data comprehensively affects the throughput, which makes the conventional ABR algorithms prohibitively complex to describe these relations in analytical expressions. This problem motivates us to exploit neural networks without relying on preconfigured analytical expressions. Through the training process, the neural networks can gradually learn the experience to cope with the irregular phenomena and extract effective features from these complex relations to improve forecast capability.

\vspace{0.2cm}
\section{Learning Sensor-Augmented Algorithms}\label{sec:training}

SA-ABR generates ABR algorithms based on the DRL model and LSTM network. The objective of SA-ABR is to maximize the expected rewards of user feedback over the whole video. Thus, instead of simply using neural networks to emphasize each temporal step of video playback, DRL enables SA-ABR to focus on the overall performance and generates the optimal bitrate selection policy over the whole video sequence. Furthermore, SA-ABR can take advantages of DRL to essentially improve the bitrate selection mechanism by forcing the agent to automatically learn better strategies without manual configuration about the throughput traces and video states.

\begin{figure}[t]
	\centering
	\includegraphics[width=0.49\textwidth]{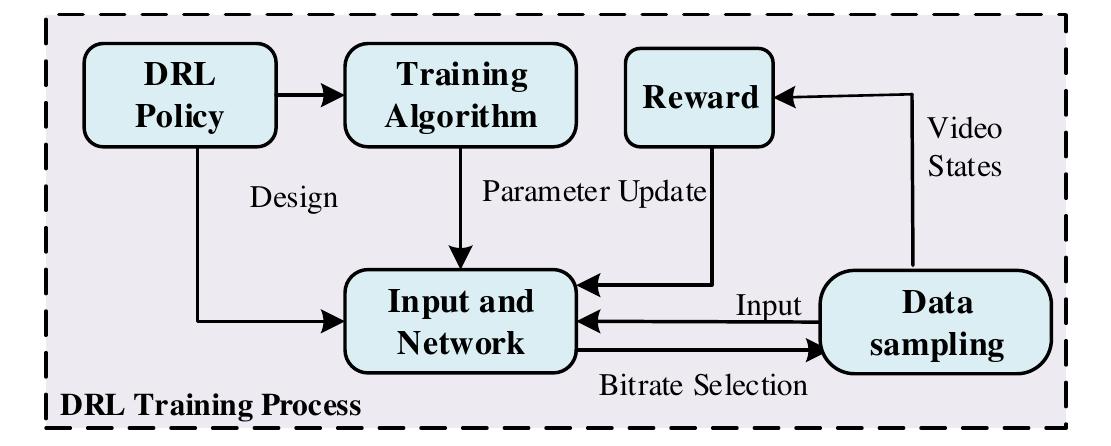}
	\caption{DRL training process.}\label{workflow}
\end{figure}

We first design a training methodology to faithfully model the dynamics of video streaming in client applications, which accelerates the training process. Then, we introduce the quantization-based preprocessing performed on sensor data before directly feeding it to the networks. Finally, we characterize the DRL training process in various aspects. As shown in Fig.~\ref{workflow}, the training algorithm and the networks are established based on the DRL policy. The networks receive a variety of inputs (e.g., the video states, the sensor data) through data sampling process and output bitrate selections for future chunks. The reward is an assessment of video quality, which motivates the network parameters constantly updated to achieve better video quality. The DRL training process is implemented as a prior task and our ABR model is running on the ground client.

 
\subsection{Training Methodology }
The first step is to design a training methodology that faithfully models the dynamics of video streaming. In fact, using actual video clients for training need to employ a video player to continuously download videos and observe video state changes. This process is very time-consuming since the download time is long and the model cannot be updated until all video chunks are downloaded. 
For example, the video in our experiment is set to have 41 chunks, with each chunk lasting two seconds. That is to say, the model needs to wait tens of seconds before being updated. 


In order to save the video download time and accelerate the training process, our training methodology directly calculates the new video states (including the buffer occupancy and the rebuffering time) based on current information. The buffer occupancy, shown as the progress bar of video players, indicates how long the video can continue to play.
Moreover, video transmissions from UAVs are mainly live streaming, which may have a time-to-live requirement of 15-45~s. This bounds the buffer level that video players can build~\cite{Akhtar:2018:OAV:3230543.3230558}. Thus, we set the maximum buffer boundary and the maximum rebuffering time to 20~s. Once the buffer level or rebuffering time exceeds this boundary, our model will be punished and obtain a very low QoE reward. Additionally, there exists a half of the round-trip-time (RTT) delay derived from the video requests sent by the client to the UAV. However, the delay is below the millisecond level which has been proved to have minimal impact on the chunk-level ABR systems~\cite{Mao2017Neural}. 

At the end of each video chunk $ t $'s download, the update of video states can be divided into two cases. First, if the download time $ f_t $ of the video chunk $ t $ is less than the beginning buffer value $ b_t$ $(b_t < 20~s) $, no rebuffering will happen. We update the new buffer size $ b_{t+1} $ by subtracting the download time $ f_t $, and then adding a two-second duration of one chunk $ t_{chunk} $. The updated buffer size $ b_t $ and the rebuffering time $ T_t $ can be expressed as
\begin{equation}\label{buffer1}
\begin{cases}
b_{t+1} = b_t+t_{chunk}-f_t, \\
T_t = 0,
\end{cases}
if \quad b_t >= f_t.
\end{equation}

Therein, the download time $ f_t $ solely relies on the video chunk's bitrate selection $l_t$ and network throughput traces  $x_{trace, t}$. The throughput traces $x_{trace, t}$ employed are collected in advance by keeping track of the wireless channel quality of UAVs in the wild. The available video bitrates in our experiment are \{300, 750, 1850, 2850\}~Kbps that correspond to video types of \{240, 360, 720, 1080\}~p.

For the second case where the download time $ f_t $ of the current video chunk exceeds the beginning buffer value $ b_t $, the rebuffering occurs. In these scenarios, our model is configured to wait for 500~ms before retrying to request the playback of next video chunk. In addition, the video chunks cannot be played until they are downloaded. Thus, the buffer size and rebuffering time are updated by

\begin{equation}\label{buffer2}
\begin{cases}
b_{t+1} = t_{chunk},\\
T_t = \left \lceil \frac{f_t-b_t}{0.5} \right \rceil \times0.5,
\end{cases}
if \quad b_t < f_t.
\end{equation}

After each chunk is downloaded, SA-ABR passes new observations of video states, including the buffer size $ b_{t+1} $ and rebuffering time $T_t$, to the RL agent. The agent then assesses the video QoE and obtains the reward to periodically update the policy. Guided by the policy, the agent makes the next bitrate decisions based on the received states. Then, new video states are regenerated by Eq.~\eqref{buffer1}-\eqref{buffer2} for the next round of update.




\subsection{Preprocessing sensor data}
Recall that the general relationships between the inherent sensor data and the throughput are summarized in Section~\ref{sec:mobility}. We incorporate the sensor data into SA-ABR to provide hints about channel variance patterns. However, the raw sensor data (such as the acceleration patterns, shown in Fig.~\ref{acce_zong}) keeps fluctuating. These noises, incurred by UAV vibrations, have little impact on throughput, while the resulting uncertainty in sensor data can interfere with the model's prediction of future throughput. Moreover, when the link distance is relatively short, the path loss is dominated by the constructive and destructive interference caused by the multipath effect~\cite{chowdhery2018aerial}. That means, the distance increment within a closer area has little impact on throughput (shown in Fig.~\ref{2a}). However, such a distance change may interfere with the model's throughput forecast. Therefore, in order to ensure the full use of the sensor data that is indicative of throughput dynamics while eliminating these noises and disturbances, we quantize the sensor data according to the degrees of its impact on throughput, instead of directly feeding it to the neural network.

The quantization scheme is designed as follows. Recall that the available video bitrates are $\{$300, 750, 1850, 2850$\}$~Kbps. Thus, the throughput for video transmission can be roughly divided into $<$0.75~Mbps, 0.75-1.85~Mbps, 1.85-2.85~Mbps and $>$2.85~Mbps, each of which can satisfy the corresponding video bitrate, with an average interval of 0.95~Mbps. That is to say, when the throughput drops by 0.95~Mbps, the optimal bitrate selection may change. Then, we sort the throughput in ascending order of the corresponding sensor data and move a sliding window to obtain the average throughput. The experimental data is from Section~II.C and the UAV is allowed to fly on random paths at arbitrary speeds. When the average throughput drops by 0.95~Mbps, the corresponding sensor value is set as a quantization threshold. This quantization process is for the overall value of the sensor data, and it is not necessary to consider the partitions of the 3D state space.
After traversing the entire experimental data corpus, we can obtain the quantization results as follows.


\begin{itemize}
	
	\item\textbf{Preprocessing distance data.} We encode distances over 50~m into ``1'', and distances within 50~m into ``0''.

	\item\textbf{Preprocessing velocity data.} The first level ``0'' represents velocity below 8~m/s, and the second level ``1'' indicates velocity within 8-12~m/s, while the third ``2'' is velocity over 12~m/s.

	\item\textbf{Preprocessing acceleration data.} The acceleration data which exceeds 18~$\text{m/s}^2$ is marked as ``1'', while the other is marked as ``0''. 
\end{itemize}


\begin{figure}[t]
	\centering
	\includegraphics[width=0.49\textwidth]{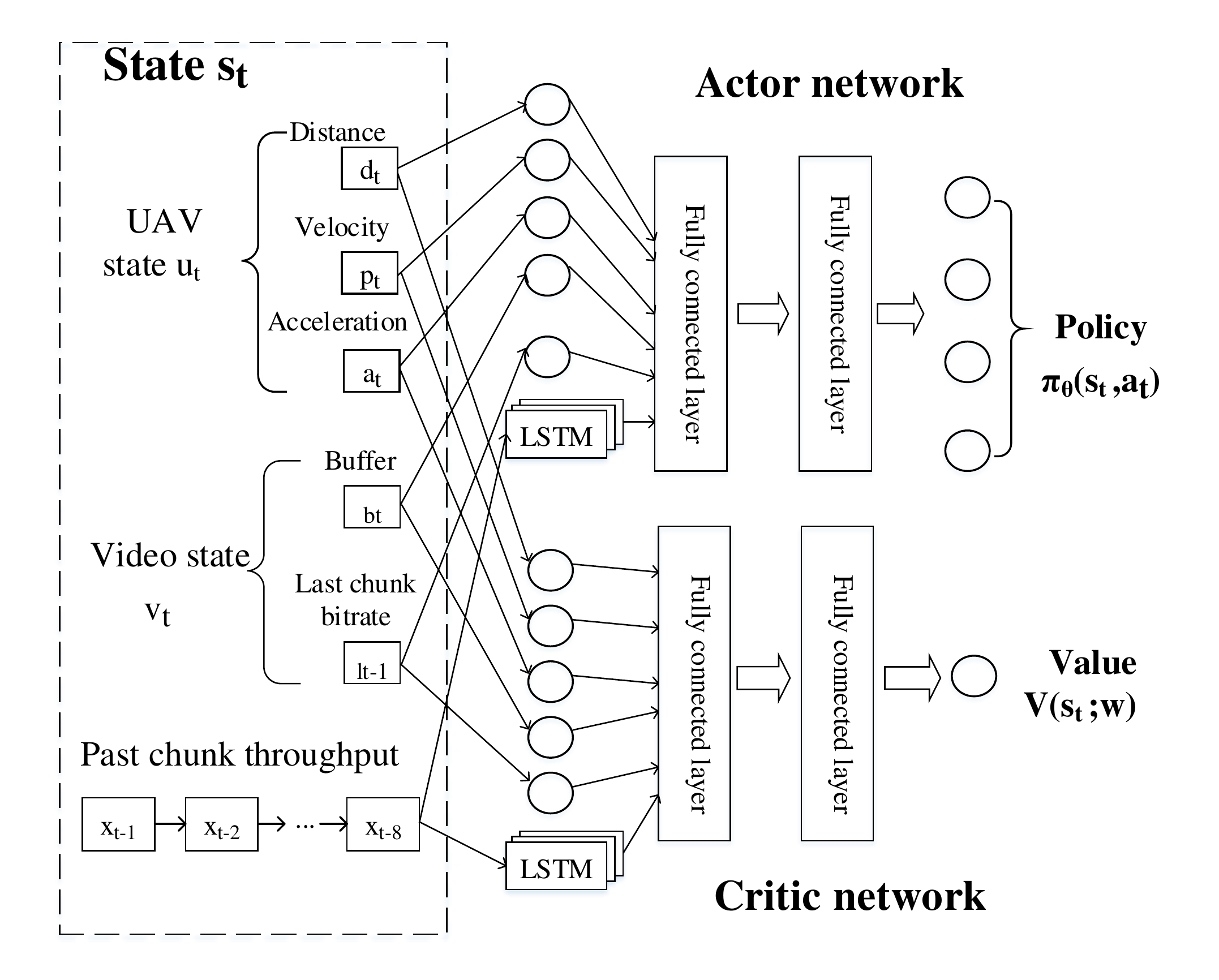}
	\caption{The network architecture designed for UAV video transmissions.}\label{network}
\end{figure}

\subsection{DRL training process}
We use the Advantage Actor-Critic~\cite{mnih2016asynchronous,Sutton1999Policy} method to generate ABR algorithms with the assistance of sensor data and LSTM networks. The training algorithm is based on on-policy policy. Although both generated by DRL algorithm, SA-ABR is very different from Pensieve~\cite{Mao2017Neural}. SA-ABR first incorporates the sensor data as input to accommodate the changes in the UAV channel. Then, it quantizes the sensor data to eliminate the noise and disturbance in the UAV channel. Moreover, it also explores the role of the LSTM network in analyzing past throughput sequences and predicting future changes. To summarize, SA-ABR takes into account the flight states of UAVs and is optimized to improve QoE in UAV video streaming, which are brand new compared to existing DRL-based solutions, including Pensieve~\cite{Mao2017Neural}.

\textbf{Input and network.} We take the current state $\vec{s}_t$ as input to the two neural networks. As shown in Fig.~\ref{network}, $\vec{s}_t$ is defined as $\vec{s}_t \triangleq (\vec{u}_t,\vec{v}_t,\vec{x}_t) = (d_t,p_t,a_t,b_t,l_{t-1},\vec{x}_t)$.
The subscript $t$ indicates that the model has finished downloading chunk $t-1$ and intends to choose chunk $t$.
The UAV's state $\vec{u}_t$ refers to its motion characteristics, obtained at the beginning of each video selection, which consists of distance $d_t$, velocity $v_t$ and acceleration data $a_t$. The video playback state $\vec{v}_t$ consists of the buffer size $b_t$ and last bitrate selection $l_{t-1}$. The vector $\vec{x}_t$ represents the average throughput of past eight chunks. In short, our input consists of five parameters and a $ 1\times8 $ vector. The typical input dimensions of RL-based ABR algorithms~\cite{Mao2017Neural,Claeys2013Design,Chiariotti2016Online,Claeys2014Design,Hooft2015A,Huang:2018:QVQ:3240508.3240545} range from two parameters~\cite{Hooft2015A} to $25 \times64\times36$ matrices~\cite{Huang:2018:QVQ:3240508.3240545}.

As shown in Fig.~\ref{network}, the past throughput data \{$x_{t-8},..., x_{t-1}$\} is fed into an LSTM network in time order for feature extraction. We choose the LSTM network because it can make full use of the temporal characteristics in the throughput sequence. That means, the earlier the throughput, the smaller its impact on the throughput forecast. This advantage improves the forecast capability and prompts us to choose the LSTM network instead of convolutional neural network (CNN).
Then, the feature vectors obtained from the LSTM network are concatenated with other state information and finally fed into the hidden layers. Although using the same structure, the actor network and the critic network are separate and have different output. 


\textbf{Deep reinforcement learning policy.} Almost all RL problems can be formulated as Markov decision processes (MDPs)~\cite{zhou2016mdash,Mao2016Resource} to achieve the optimal solutions. 
Generally, the whole MDP consists of a state set $ S $, an action set $ A $ and a reward set $ R $, each of which can be expressed as a sequence tuple $ M = \{s(t), a(t), r(t), s(t+1)\} $. In our system, $ s(t) $ denotes the current state at video chunk $ t $ and $ a(t) $ indicates the selected video bitrate based on the current state. The reward $ r(t) $ entirely depends on the states and model's reactions and is expressed as $ r(s(t), a(t)) $. In the context of video streaming, each time a video chunk is downloaded, SA-ABR obtains a reward to evaluate the current state and action. The discounted cumulative reward can be expressed as
\begin{equation}
R_t = r_t+\gamma r_{t+1}+\gamma^2 r_{t+2}+\gamma^3 r_{t+3}+\cdots,
\end{equation}
where $\gamma \in [0,1]$ denotes the discounted factor, and $R_t$ represents the discounted cumulative reward from time chunk $ t $ to the end. 

As shown in Fig.~\ref{network}, two neural networks are included in our proposed model. The goal of actor network is to find a strategy $\pi:\pi_\theta(s,a)\rightarrow[0,1]$ to maximize the total reward. In our system, $\pi_\theta(s,a)$ is the probability distribution over different video bitrate choices. With this distribution, each video bitrate is selected based on its probability via a stochastic policy, i.e., the one with the highest probability is the most likely to be picked up. The duty of critic network is to make an objective assessment $V(s_t; w)$ for the current state $s_t$. 

Nevertheless, SA-ABR does not directly increase the discounted cumulative reward $R_t$ as the update direction~\cite{Sutton1999Policy}. Instead,
$R_t$ is subtracted by a baseline $b_t$, 
and $R_t-b_t$ can be replaced by the advantage function $A^{\pi_\theta}(s_t,a_t) = Q^{\pi_\theta}(s_t,a_t) - V(s_t; w)$ in the network. This represents the difference in the cumulative reward between the expected value and the actual value after selecting the action $a_t$ based on policy $\pi_\theta$ at $s_t$.

\textbf{Training algorithm.} 
The key step in the actor network is to calculate the advantage function $A^{\pi_\theta}(s_t,a_t)$. In our system, we use the n-step Temporal-Difference (TD) method~\cite{Konda2003On} to calculate the advantage function in the actor network, which is given as
\begin{equation}
	Q^{\pi_\theta}(s_t,a_t) = \sum_{k=0}^{k=n-1} \gamma^k r_{t+k} +\gamma^n V(s_{t+n}; w),
\end{equation}
where $\gamma \in [0,1]$ denotes the discounted factor. $Q^{\pi_\theta}(s_t,a_t)$ is not simply the discounted cumulative reward from chunk $ t $. Instead, it sets the final reward as $V(s_{t+n}; w)$, which is the assessment from the critic network of chunk $ t+n $. In our experiment, the subscript $ t+n $ refers to the end chunk in a video. By performing the subtracting operation, we can finally obtain the advantage function $ A^{\pi_\theta}(s_t,a_t) $. 

At the training phase, the goal of the actor network is to maximize the advantage function, i.e., making better decisions than the current policy. Thus, the parameter of the actor network $\theta$ is updated via a stochastic gradient ascent algorithm
\begin{equation}\label{actor}
	\theta \leftarrow \theta + \alpha \sum_t\nabla_\theta\log\pi_\theta(s_t,a_t)A^{\pi_\theta}(s_t,a_t),
\end{equation}
where $\alpha $ is the learning rate. $ \nabla_\theta\log\pi_\theta(s_t,a_t)$ shows how to change the parameter $ \theta $ in order to achieve the goal. Additionally, in order to improve the generalization capability of the network, SA-ABR applies the dropout technique to reduce overfitting and add a regularization term to the update of the actor network. This term is the entropy of the probabilities over bitrate selections $ H(\pi_\theta(\cdot |s_t))$, which encourages exploration and prevents overfitting.

For the critic network, it aims at making an accurate assessment for all the states of experiments during training. We use the standard TD method to calculate the loss function of the critic network and minimize the value. Therefore, we can update the parameter of the critic network $w$ through a stochastic gradient descent algorithm
\begin{equation}\label{critic}
	w \leftarrow w-\alpha' \sum_t\nabla_w(r_t+\gamma V(s_{t+1}; w)-V(s_{t}; w))^2,
\end{equation}
where $\alpha'$ is the learning rate, $V(s_{t}; w)$ and $V(s_{t+1}; w)$ are respective assessments from the critic network of chunk $t$ and chunk $t+1$.

\textbf{Data sampling.} 
Based on the on-policy learning algorithm, the RL agent is updated periodically as data arrives, and then follows the updated strategy to sample the new data. To reduce the correlation between the data sampled from one agent and accelerate the training process, we run ten agents in parallel to experience different states, transitions and environments. These elements form a minibatch of $ \{s_t,a_t,r_t,s_{t+1}\}$ tuples and are sent back to the central agent. The central agent then uses the actor-critic algorithm to compute the policy gradient (Eq.~\eqref{actor} and Eq.~\eqref{critic}) and updates the networks. Note that this algorithm does not require replay memory (e.g., Deep Q-Network (DQN)) and the extreme version can be directly trained on video clients to adapt to the varying UAV conditions, which shows the advantages of using actor-critic algorithm to adapt to the UAV video transmission.


\textbf{Reward.} Reward $r$ is given after each chunk is downloaded. It reflects the performance of each bitrate selection according to whether the video quality meets the requirements of viewers. In our system, we consider a general QoE metric~\cite{Yin2015A,Mao2017Neural} as a reward judging criterion, which is defined as

\begin{equation}\label{QoE_definition}
QoE = q(l_t)-\mu T_t-|q(l_t)-q(l_{t-1})|,
\end{equation}
where $q(l_t)$ represents the user perception for video bitrate $l_t$, $T_t$ the rebuffering time and $ |q(l_t)-q(l_{t-1})| $ the jitters between video chunks. Thus, the QoE is determined by three factors including the bitrate utility $q(l_t)$, the rebuffering penalty $\mu T_t$ and the smoothness penalty $ |q(l_t)-q(l_{t-1})| $.

Generally, there are several definitions of $q(l_t)$~\cite{Yin2015A,Spiteri2016BOLA,Mao2017Neural} in QoE metrics. We use the following definition~\cite{Spiteri2016BOLA,Mao2017Neural}
\begin{equation}
QoE_{log} : q(l_t) = \log(l_t/l_{min}).
\end{equation}
The $QoE_{log}$ is chosen because this kind of metric does not pursue excessive clarity, for the increase in reward gradually shrinks when switching to high bitrate selection. Therefore, it is more practical to decline the rebuffering time and improve the smoothness to maximize the reward. This preference is also very suitable in the video transmission scenario, since UAVs are often used for real-time video capturing and transmission. 

\vspace{0.2cm}
\section{Implementation}\label{sec:implementation}
This section encompasses three aspects. To begin with, we characterize the neural network architecture and all the hyperparameter settings during the training process. Next, we give a brief introduction of our UAV platform. Finally, we elaborate on the collected network traces used in our experiment.

\textbf{Neural network architecture.} In this part, we elaborate on the specific design of our neural network architecture and all the hyperparameters in the experiment. First, we feed a sequence of past eight chunks' throughput in time order to an LSTM network, which is constructed with two-layer LSTM cells, with 64 hidden unites each. Each input contains one throughput in the past and the step size is one. Then, the resulting vector from this LSTM network is flattened and combined with other inputs including video playback states and UAV's flight states before being imported to two fully connected layers, with 30 and 10 hidden units, respectively. The actor network and the critic network have the same input and structure, while they are different in network parameters and output. We add a softmax layer and obtain a probability distribution of the bitrate selections for the actor network, while setting the state evaluation as output for the critic network. Furthermore, the input sizes of both networks are 13. Within the training period, we set the discount factor $\gamma$ to 0.99, and configure the learning rates $\alpha$ and $\alpha'$ to $3\times10^{-5}$ and $1\times10^{-2}$, respectively. Additionally, the reward factor $\mu$ is set to 2.26 according to the QoE metric we choose. 

\begin{figure}[t]
	\centering
	\includegraphics[width=0.495\textwidth]{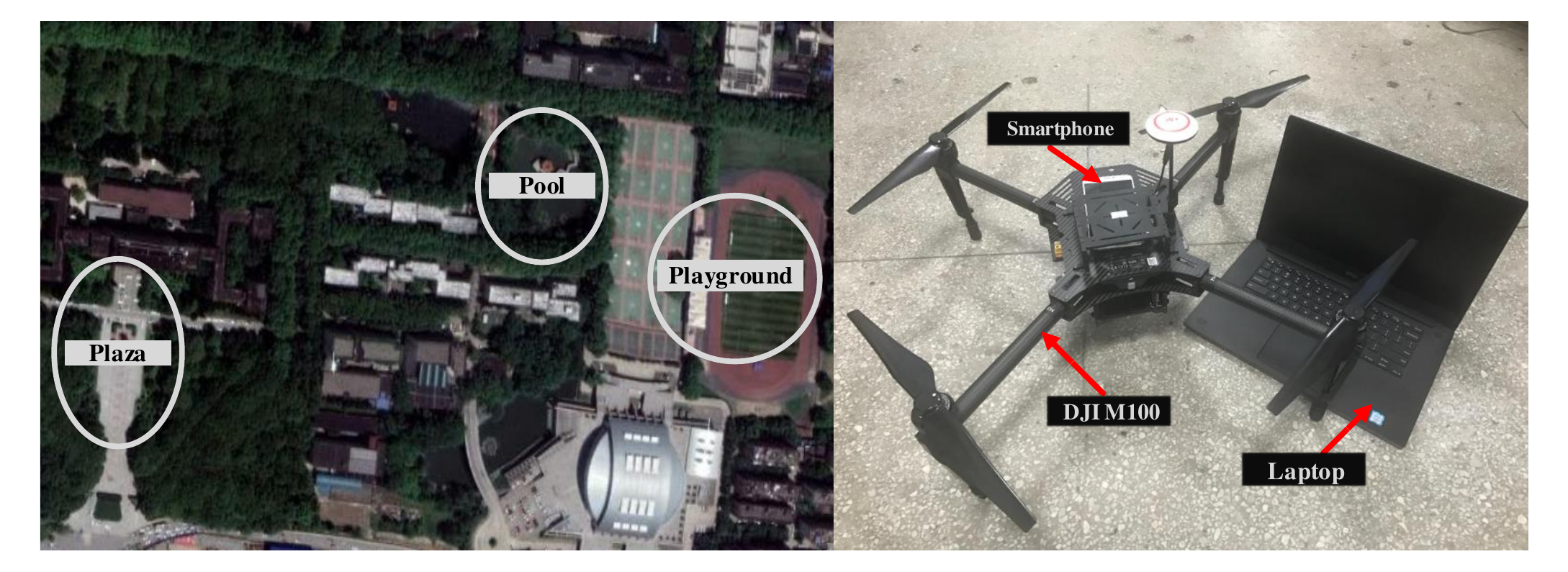}
	\caption{The sites of our experiments and the experimental platform. The white circles in the left picture denote the flying areas of our experiments including a playground, a plaza and a pool. The right part is a DJI Matrice 100 UAV platform. }\label{uav}
\end{figure}

\textbf{Hardware setup.} As shown on the right part of Fig.~\ref{uav}, we build a wireless link between a UAV and a laptop on the ground through the IEEE 802.11n protocol. SA-ABR is a client-based model and the key ABR algorithm runs on the laptop. An Android smartphone is attached on our UAV platform, DJI Matrice 100, and transmits the video and sensor data to the laptop on the ground. Additionally, the data file is programmed to be transmitted through a WiFi channel, which is different from the control channel of the UAV to avoid interference. We use the 2.4~GHz band for transmission with a channel bandwidth of 20~MHz. During the transmission process, the laptop on the ground calculates the number of TCP packets in the application layer per second to obtain the average throughput, which tracks the wireless-link dynamics. Another function of the laptop is to emulating the video playback states, which we specifically describe in Section~\ref{sec:training}. Based on these available messages, SA-ABR can gradually adapt to the throughput dynamics through the training process. 



\textbf{Network traces.} In order to model the real-world wireless channel condition on UAV, we collect the throughput data by flying the UAV in the wild. Since the available video bitrates in our experiment are \{300, 750, 1850, 2850\}~Kbps, the throughput is multiplied by weight to match the video quality, which is common in the scenes where the UAV transmits videos to multiple clients. The experimental sites includes a playground, a plaza and a pool. The flight trajectories are randomly distributed in these areas and the flying velocities range from 0 to 19.5~m/s. The height of the UAV is set at around 25~m for safety. We record a total of 1000 throughput traces, with each trace spanning 100~s. In our experiment, we use a random sample of 80$ \% $ in the data corpus as the training set, while the remainder 20$ \% $ is used as the test set. The throughput ranges from 0 to 20~Mbps. 

\vspace{0.2cm}
\section{Evaluation}\label{sec:evaluation}

In this section, we conduct a series of experiments to evaluate SA-ABR from three aspects as described below.

\begin{figure*}[t]
	\centering
	\includegraphics[width=1\textwidth]{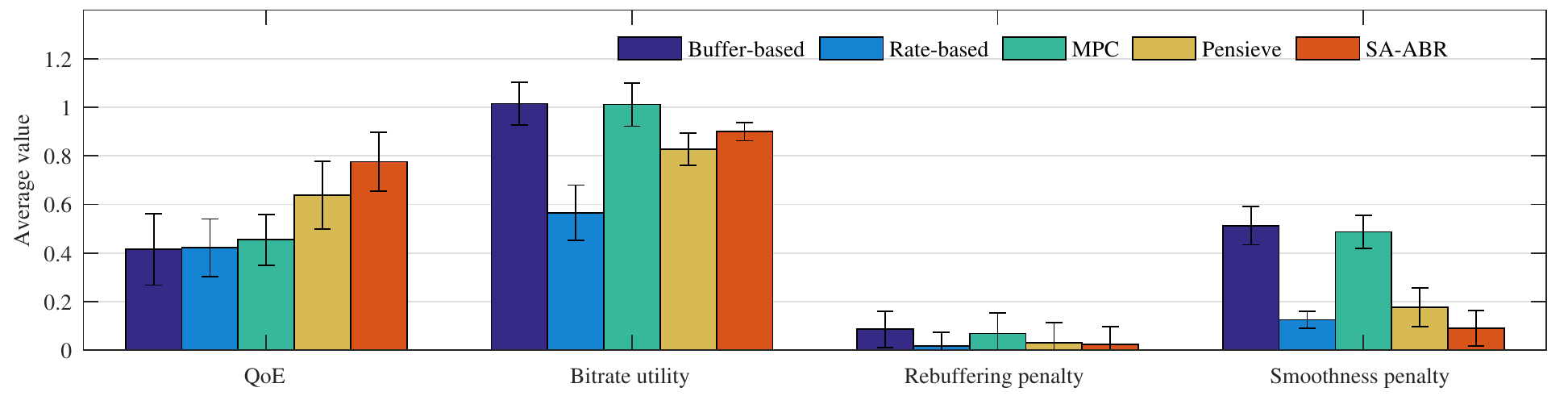}
	\caption{Comparing SA-ABR with a variety of existing ABR algorithms by not only presenting the average QoE value for one chunk, but also analyzing their respective performance on each individual component of our considered QoE metric (presented in Section~\ref{sec:training}). }\label{reward_zong}
\end{figure*}

\begin{figure*}
	\centering
	\subfigure[The corresponding CDF picture of the average QoE value]{
		\includegraphics[width=0.23\linewidth]{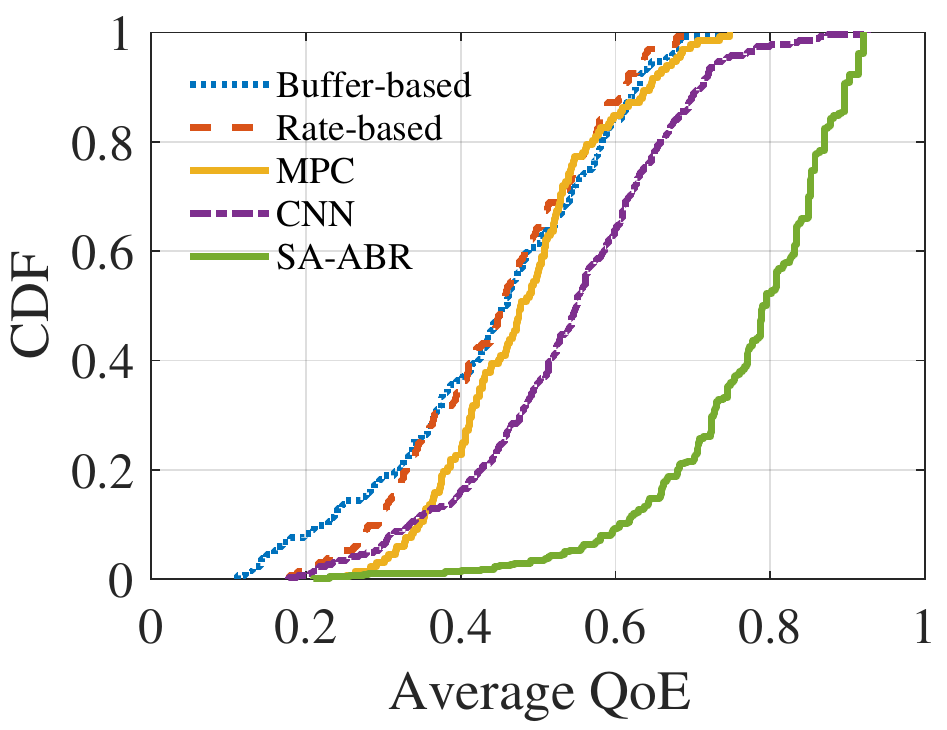}\label{fen_a}}
	\hfill
	\subfigure[The corresponding CDF picture of the average bitrate utility]{
		\includegraphics[width=0.235\linewidth]{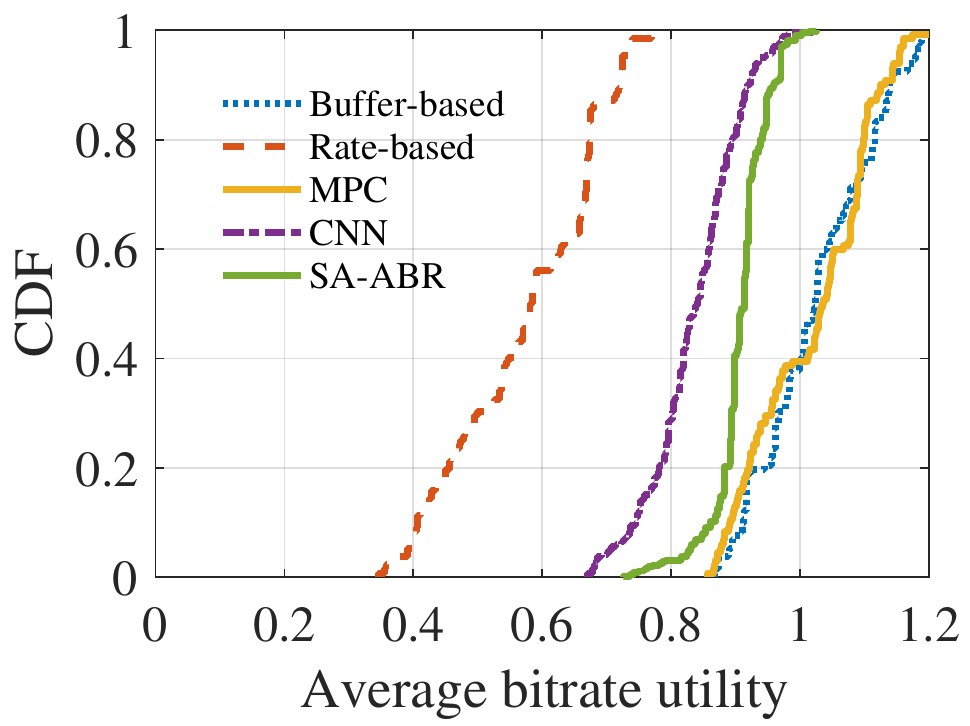}\label{fen_b}}
	\hfill
	\subfigure[The corresponding CDF picture of the average rebuffering penalty]{
		\includegraphics[width=0.23\linewidth]{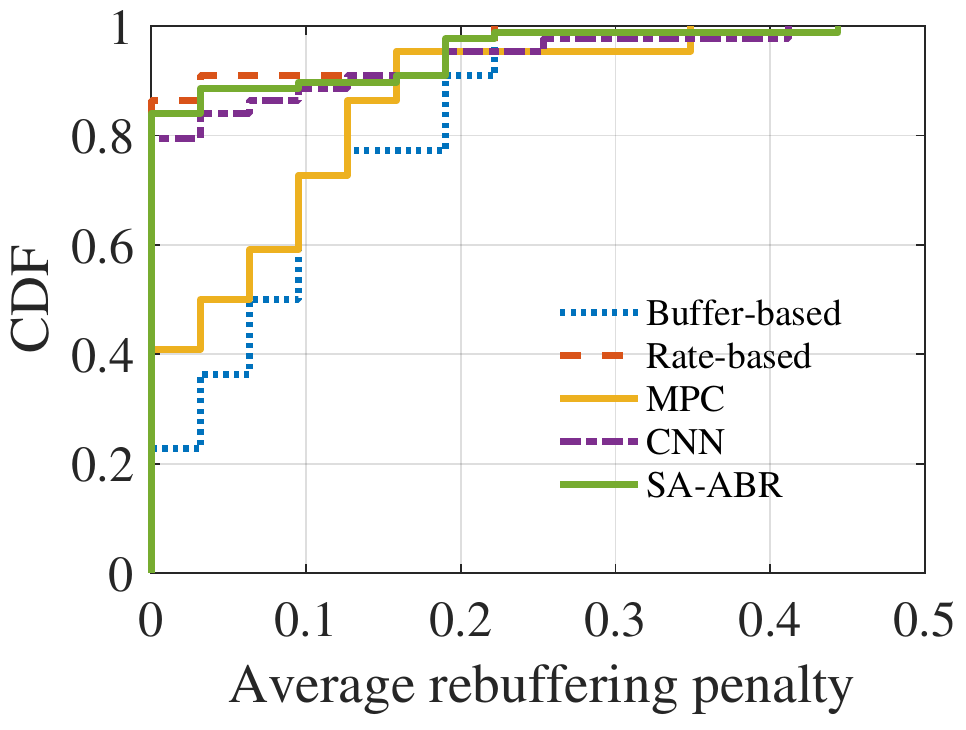}\label{fen_c}}
	\hfill
	\subfigure[The corresponding CDF picture of the average smoothness penalty]{
		\includegraphics[width=0.23\linewidth]{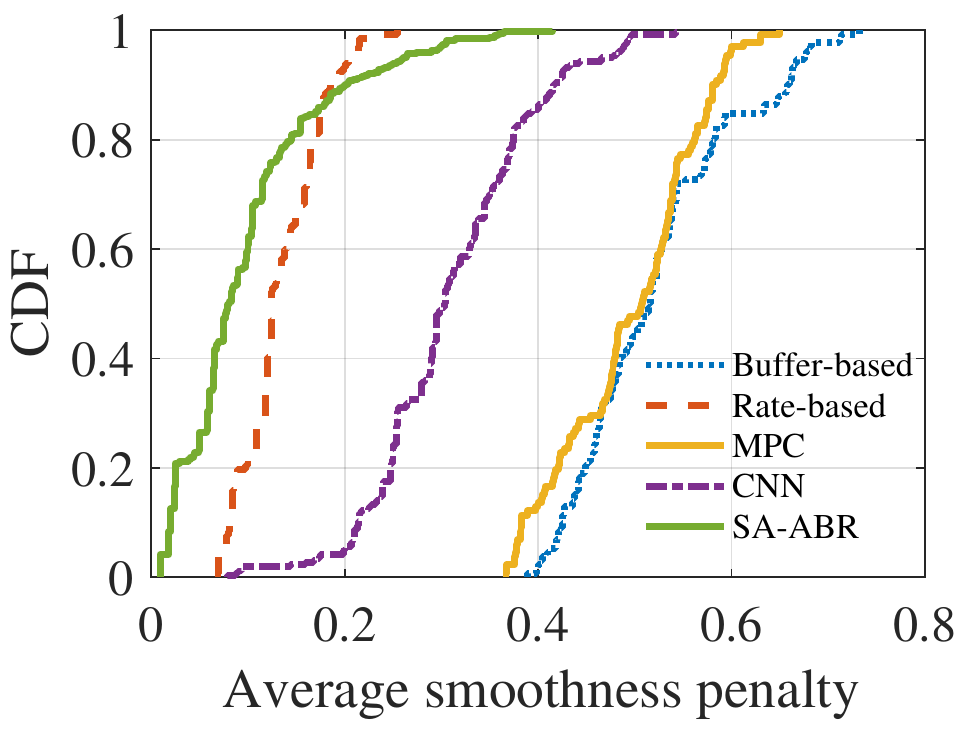}\label{fen_d}}
	\\
	\caption{The results of comparisons between SA-ABR and other existing ABR algorithms in the form of CDF.}
	\label{reward_fen} 
\end{figure*}

\begin{itemize}
	\item \textit{How does SA-ABR compare with state-of-the-art ABR algorithms in terms of video QoE?} We test several ABR schemes including kinds of fixed control rules and RL algorithms, and then perform comparative experiments with SA-ABR. The result shows that SA-ABR always performs better compared with other algorithms and is able to outperform the best ABR schemes by 21.4$ \% $ in terms of average QoE value (Figs.~\ref{reward_zong}, \ref{reward_fen}, and \ref{action_buffer}).
	
	\item \textit{Does SA-ABR benefit from LSTM network?} In the framework design of SA-ABR, we exploit the LSTM network to extract the time-series features from the past throughput experience. To specifically evaluate the advantage, we propose a baseline model that is similar to our network architecture, but replacing the LSTM network with CNN. By comparing these two models, we find SA-ABR still presents its improvements in average QoE (Fig.~\ref{reward_lstm}).
	
	\item \textit{What is the advantage of feeding various sensor data into the neural network and can the network effectively filter out the confusing information and extract useful features from it?} We conduct an experiment to compare SA-ABR with the same network architecture without sensor assistance. Results in Fig.~\ref{reward_sensor} show the performance improvement for SA-ABR with the assistance of sensor data.
\end{itemize}

\subsection{SA-ABR vs. Existing ABR Algorithms}
We compare SA-ABR with a variety of existing state-of-the-art algorithms which generate the ABR algorithm in completely different ways including fixed control rules and RL. These algorithms perform bitrate adaptations mainly based on the past throughput experience and video playback states without the assistance of sensor data. The detailed principles of these algorithms are illustrated below.

\begin{itemize}
	\item \textbf{Buffer-based policy}~\cite{Huang2014A} is an algorithm that chooses the video bitrate only based on the playback buffer occupancy. The goal is to reach balanced states that ensure the avoidance of unnecessary rebuffering while maximizing the average video bitrate. The model is manually configured to keep the buffer occupancy above five seconds and when the buffer occupancy exceeds 15~s, the highest available bitrate is automatically selected. 
	
	\item \textbf{Rate-based policy}~\cite{Jiang2012Improving} exploits the harmonic bandwidth estimator to compute the harmonic mean of the last five throughput samples, which provide robust bandwidth estimates for future chunks. Thus, this model can automatically select the highest bitrate that does not exceed the expected channel capacity.
	
	\item \textbf{MPC}~\cite{Yin2015A} proposes a concrete ABR workflow that can optimally combine the advantages of future throughput prediction and buffer-based functions. The algorithm uses the same approach as Rate-based models to provide throughput forecasts, based on the past throughput trajectory. Then, MPC can map the collected information including the throughput prediction value, previous bitrate and buffer occupancy to future bitrate selections of video chunks.
		
	\item \textbf{Pensieve}~\cite{Mao2017Neural}, which is based on the RL algorithm, has been experimented to use in the video bitrate adaptation subject with no pre-programmed control rules or explicit assumptions of the environments. The model uses CNN to extract effective features from the input data, including the past throughput trajectories and video states, and learns automatically through the RL algorithm to make better ABR decisions.
		
\end{itemize}

In our experiment, SA-ABR is trained to obtain the optimal policy for higher QoE metric rewards, using the training set described in Section~\ref{sec:implementation}. Although SA-ABR is generated from the limited training set, its performance can be extended to the test period in which SA-ABR can still make right decisions when encountering states that are never present in the training set. The reason is that what SA-ABR learns through DRL in the training process is not just the limited state-action pairs, but a continuous neural network function that maps the successive state space to actions.
Moreover, the parameters of the aforementioned existing ABR algorithms are also adjusted accordingly to adapt to the varying UAV channel capacity. We evaluate the performance of all the ABR algorithms based on the same test set (Section~\ref{sec:implementation}). 

In addition, we use the same QoE function (Eq.~\eqref{QoE_definition}) to assess all the ABR algorithms. Besides, three components (Eq.~\eqref{QoE_definition}) of the QoE definition, including the bitrate utility, the rebuffering penalty and the smoothness penalty, are also analyzed to better evaluate the video performance.


Fig.~\ref{reward_zong} shows the average and variance values of QoE for one video chunk. 
Note that the average QoE reward of SA-ABR is 21.4$\%$ higher than that of Pensieve, which presents as the best known ABR algorithm. 
The detailed causes of the QoE improvement are represented in the following histograms. The main reasons for the gain in the average QoE compared with Pensieve are improvements in bitrate utilization and smoothness. The average bitrate utility exceeds Pensieve by 10.8$\%$ while the smoothness penalty is reduced by 35.3$\%$. As mentioned above, the characteristics of the $ QoE_{log}$ metric lead to its slower growth in value as the bitrate increases, i.e., the reward of selecting higher bitrates (1850~Kbps and 2850~Kbps) may not counteract the rebuffering and smoothness punishment to some extent. However, keeping playing videos at lower bitrates (300~Kbps and 750~Kbps) all the time will undoubtedly affect the users' viewing experience. With the assistance of the LSTM network and sensor data, our RL-based system can better weigh the gains and losses of choosing higher video bitrates, i.e., increasing the average bitrate utility in smoother trend on the basis of no rebuffering increases. It is a step forward in ABR video streaming algorithm under the UAV channel environments. 

In addition, as illustrated in Fig.~\ref{reward_zong}, we observe that the bitrate utility of SA-ABR and Pensieve all present lower than MPC and buffer-based algorithms. 
The reason accounting for this phenomenon is that the DRL algorithm prompts SA-ABR to achieve performance gain in a more balanced way. Thus, although not achieving the maximum bitrate utility, SA-ABR has the minimum rebuffering time and the highest smoothness, which achieves a balanced state in three components of the QoE metric. In contrast, MPC and buffer-based algorithms perform poorly in decreasing rebuffering and improving smoothness. The distributions of three QoE components are also explicitly exhibited in Fig.~\ref{reward_fen} in the form of CDF, which show gaps between different ABR algorithms. Nevertheless, to further increase the average video bitrates is a constant challenge we need to overcome in the future.





\begin{figure}
	\centering
	\includegraphics[width=0.48\textwidth]{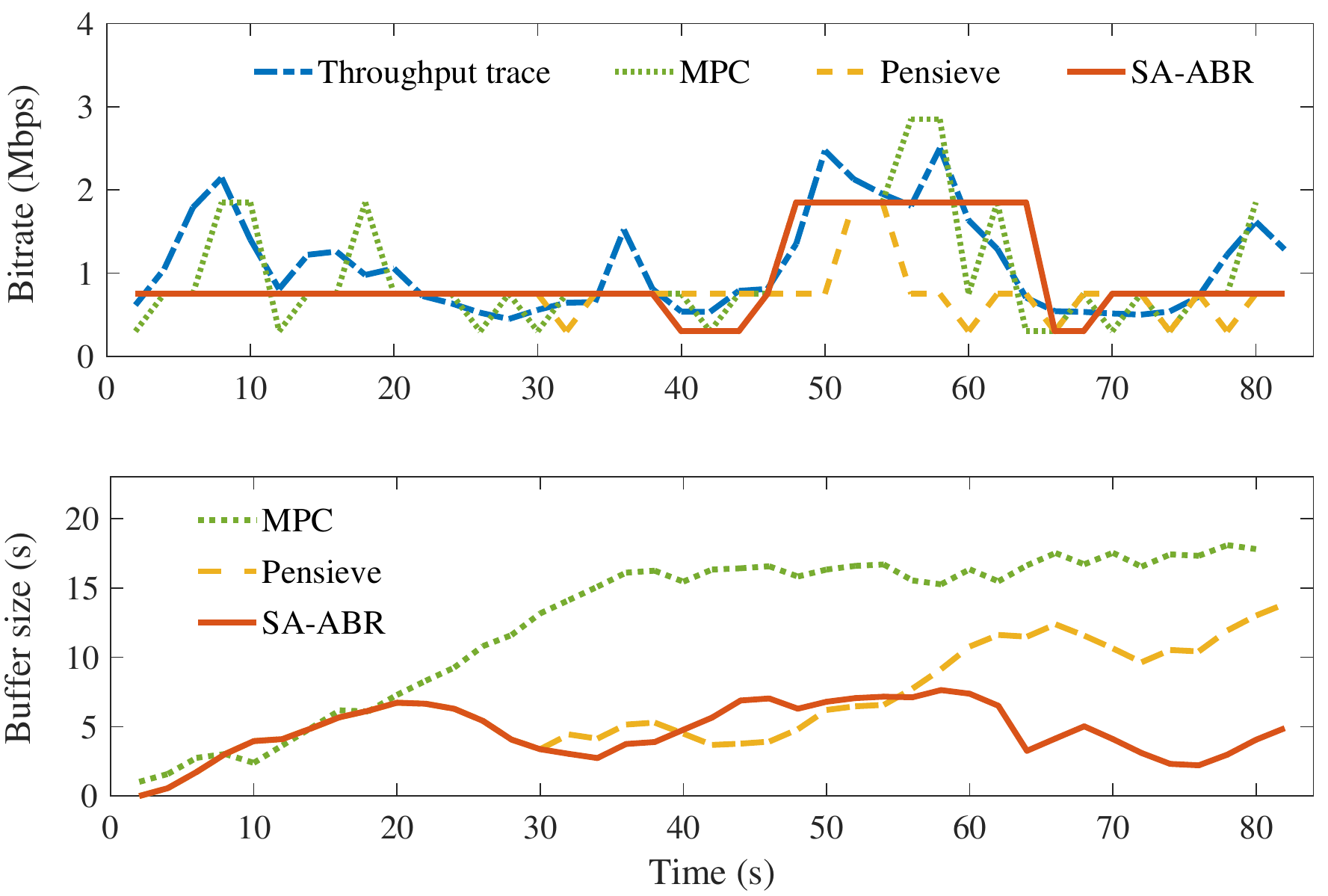}
	\caption{The upper figure is a complete network throughput trace and respective bitrate selections including MPC, Pensieve and SA-ABR over an entire video. The bottom figure is the corresponding buffer occupancy curves caused by bitrate selections above.}\label{action_buffer}
\end{figure}

Fig.~\ref{action_buffer} (top subfigure) first depicts the network throughput traces and respective bitrate selections made by  MPC, Pensieve and SA-ABR over a period of video in our data corpus. 
We can observe that MPC selects the highest bitrate (2850~Kbps) at 56~s. However, it immediately switches to a lower bitrate (750~Kbps) when the throughput begins to fall. Throughout the entire playback process, the bitrate selections of MPC fluctuate constantly with the throughput dynamics, which gives viewers bad experiences. As for Pensieve, the model does not make full use of the channel capacities and switches to a lower bitrate at 56~s due to the uncertainties in the future even though the current throughput is still very high. In contrast, SA-ABR is courageous to select the high bitrate (1850~Kbps) at 48~s and keep the bitrate for a while for video smoothness, which verifies the forecast accuracy and the ability to balance the bitrate utility and buffer occupancy.
 
Moreover, Fig.~\ref{action_buffer} (bottom subfigure) further shows the changes in buffer occupancy corresponding to each chunk' s bitrate selection for these three ABR algorithms. Compared with other algorithms, SA-ABR can make full use of the buffer occupancy, while the other models waste a lot of buffer resources.

\subsection{LSTM Network vs. CNN}

SA-ABR applies the LSTM network to extract effective time-series features from the past throughput experience, which is described in detail in Section~\ref{sec:training}. However, despite the fact that it outperforms all the listed state-of-the-art ABR algorithms, it still cannot certify whether SA-ABR benefits from the LSTM networks, as Pensieve lacks the assistance of sensor data to become the comparative experiment. Therefore, we set a baseline model which has the same architecture as SA-ABR but replaces the LSTM network with CNN.

\begin{figure}
	\centering
	\includegraphics[width=0.47\textwidth]{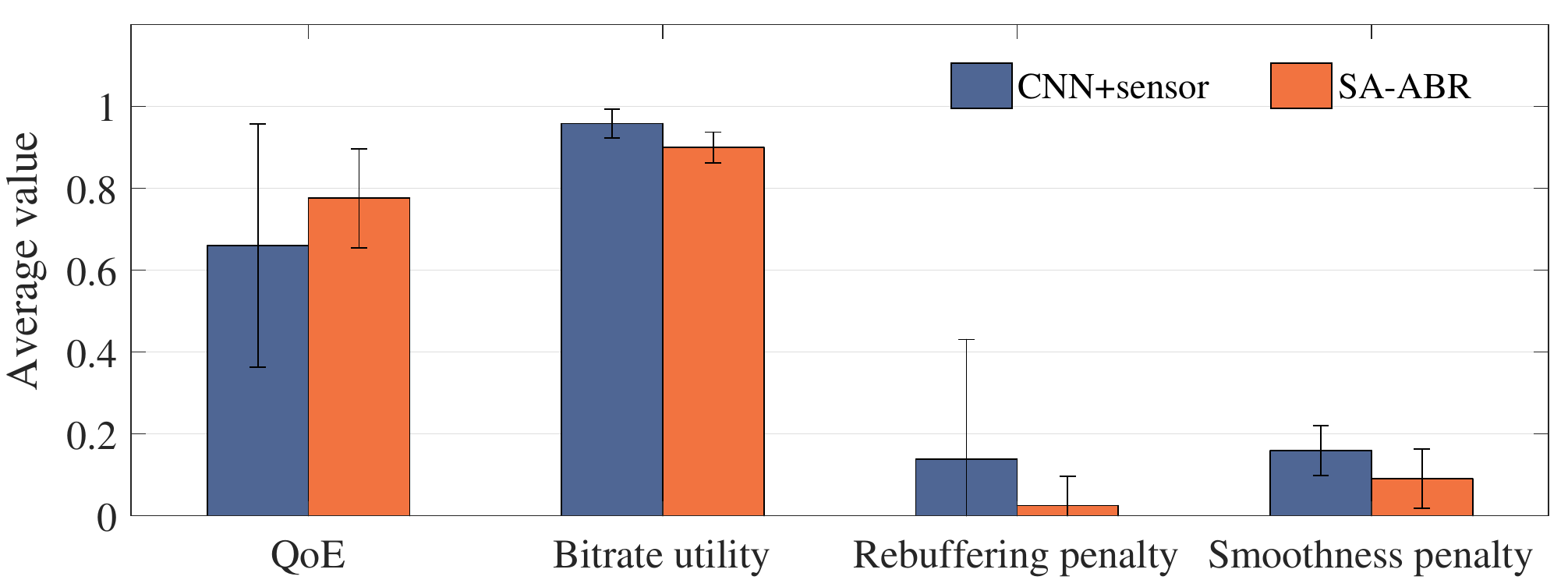}
	\caption{Comparing SA-ABR with the same architecture that doesn't exploit LSTM network. Instead, this baseline model uses CNN, also with the assistance of sensors.}\label{reward_lstm}
\end{figure}

The results of comparisons are shown in Fig.~\ref{reward_lstm}. SA-ABR's average QoE is 17.5$\%$ higher than the baseline and the gain comes from its ability to limit rebuffering and smoothness penalty. Therein, SA-ABR reduces rebuffering penalty by 88.6$\%$ through maintaining sufficient buffer occupancy to handle the risk of unpredicted fluctuations in channel capacity. This phenomenon indicates that the LSTM network masters the dramatically varying features of communication dynamics through analyzing past throughput experience. Additionally, with the assistance of the LSTM network, SA-ABR decreases the smoothness penalty by 49.4$\%$, based on the robust predictive function, 
which provides a more comfortable viewing experience. 

Moreover, to analyze how many past throughput samples are necessary to be fed to the LSTM network for better throughput forecast, we conduct an experiment to compare several versions of SA-ABR, each with a different number of throughput samples. As shown in Fig.~\ref{cdf_Raw_2816}, compared to two throughput samples, eight throughput samples enable the LSTM network to extract more temporal information from the throughput sequence, which leads to a significant gain in the average QoE reward. However, when increasing the past throughput samples to 16, the QoE gain is marginal. That means, eight throughput samples are sufficient for the throughput prediction and we select the past eight throughput samples as input to the LSTM network.

\subsection{Sensor Assistance vs. No Sensor Assistance}

For the purpose of better understanding the QoE gains obtained from the assistance of sensor data, we analyze SA-ABR's performance on individual terms of the QoE metric. To avoid interference from other related factors, such as the network type, we present another baseline model for comparison which also has the same network architecture as SA-ABR but lacks the assistance of sensor data. This baseline model is also trained to obtain the optimal policy for higher QoE reward in the experiment. 

\begin{figure}
	\centering
	\includegraphics[width=0.45\textwidth]{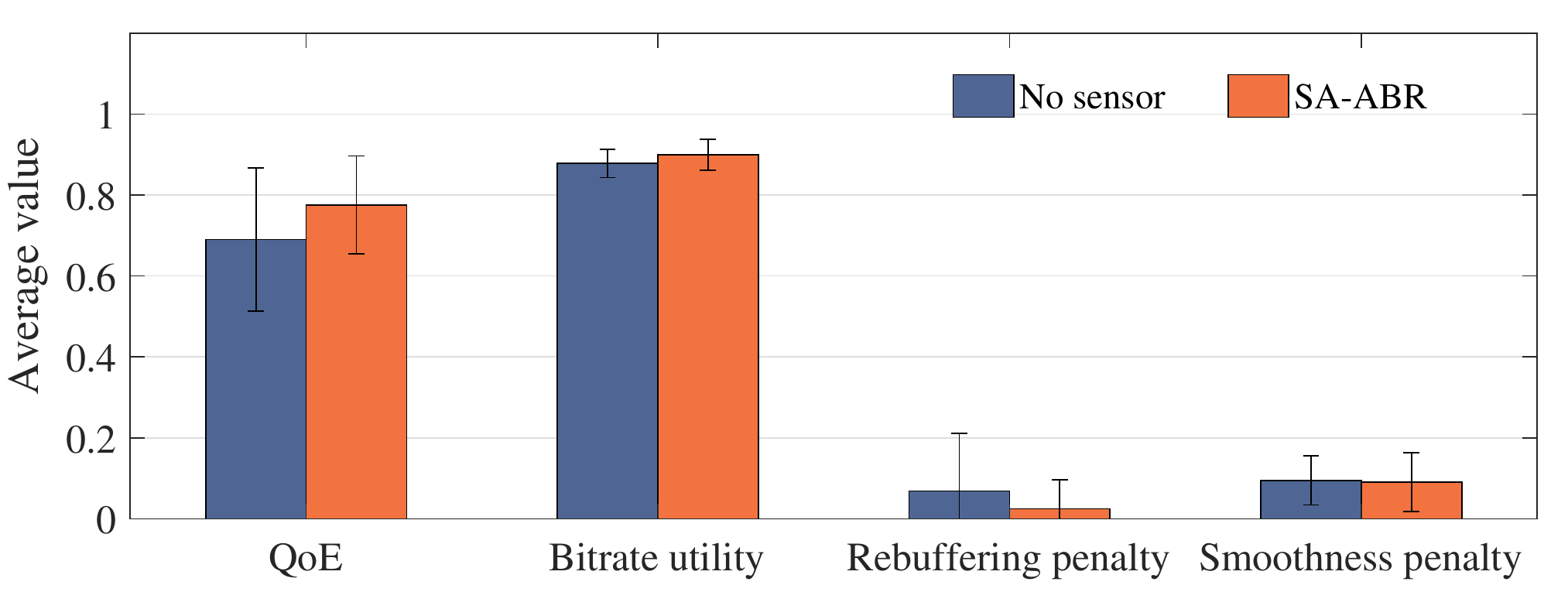}
	\caption{Comparing SA-ABR with the same architecture without the use of sensor data. }\label{reward_sensor}
\end{figure}

Specifically, Fig.~\ref{reward_sensor} shows the results of comparisons between SA-ABR and the baseline without sensor assistance. 
We observe that SA-ABR outperforms the baseline and the gain in the QoE reward ranges from 7.0$\%$ to 30.7$\%$. The reason for such a variance in the performance gain is that the improvement from the sensor data is closely connected to the flight conditions of the UAV. When the UAV flies slowly and the channel is relatively stable, the benefit of the sensor data is marginal, resulting in a lower performance gain. Whereas, when the UAV flies at a high speed, it causes unpredicted fluctuations in the UAV channel capacity. In this case, the sensor data can indicate the throughput changes and provide hints for SA-ABR to enhance the forecast capability and achieve better QoE rewards.

Other evaluation metrics such as the bitrate utility and the rebuffering penalty also verify the important role of the sensor data.
As shown in Fig.~\ref{reward_sensor}, SA-ABR increases the bitrate utility by 6.9$\%$. Although the gain is not large, it still embodies the raising abilities to challenge high-definition video chunks, resulting from the enhanced predictive accuracy. 
Moreover, SA-ABR is able to minimize the rebuffering time while ensuring the bitrate utility of video chunks, even if the channel condition is poor. As depicted in Fig.~\ref{reward_sensor}, the reduction ratio of the rebuffering penalty reaches 57.1$\%$. Furthermore, the change in the smoothness penalty is negligible, which indicates that the use of sensor data does not play a major role in maintaining sufficient smoothness. 

\begin{figure}
	\centering
	\includegraphics[width=0.41\textwidth]{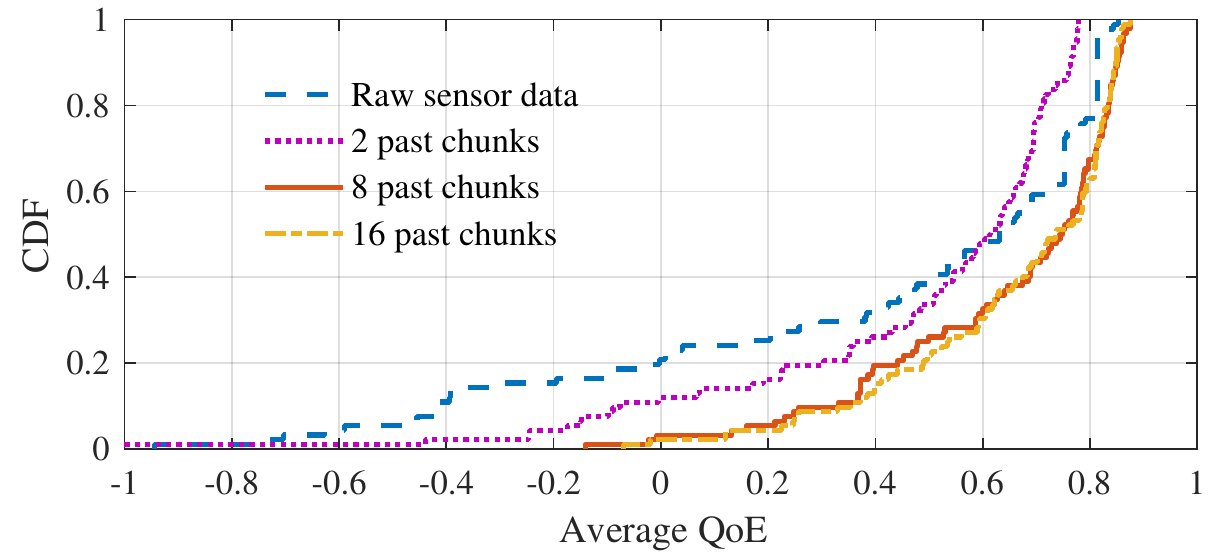}
	\caption{Comparing several versions of SA-ABR that fed by different numbers of throughput samples or raw sensor data.}\label{cdf_Raw_2816}
\end{figure}

To further verify the advantages of quantizing the sensor data, we conduct a comparative experiment in which we feed the raw sensor data into the networks. The results are shown in Fig.~\ref{cdf_Raw_2816}. Note that without the quantization process, the variance level of QoE increases while the average value declines, which indicates that the disturbance and noise from the raw sensor data prevent models from getting the optimal strategy. In other words, the quantization process eliminates the noise while ensuring  full utilization of the sensor data.

\vspace{0.2cm}
\section{Related work}\label{sec:related work}

\textbf{ABR algorithm.} The ABR algorithm essentially follows a dynamic selection mechanism that has been widely used in various fields~\cite{ChenTWLDL18,wang2016software}.
Existing ABR algorithms fall into two categories: fixed control rules~\cite{Jiang2012Improving,Sun2016CS2P,Huang2014A,Spiteri2016BOLA,Zhi2014Probe,Yin2015A,KimXMAS,zhou2016mdash,Akhtar:2018:OAV:3230543.3230558,xie2017dynamic,xu2018qoe} and learning method~\cite{Claeys2013Design,Chiariotti2016Online,Claeys2014Design,Hooft2015A,Mao2017Neural,Huang:2018:QVQ:3240508.3240545,Yeo:2018:NAC:3291168.3291216,jiang2018chameleon, sengupta2018hotdash,guo2019buffer,kan2019deep}. The majority of existing fixed control rules generate ABR algorithms based on the available bandwidth estimates (rate-based algorithms~\cite{Jiang2012Improving,Sun2016CS2P}), playback buffer occupancy (buffer-based algorithms~\cite{Huang2014A,Spiteri2016BOLA,xie2017dynamic}) or the hybrid
methods of combining these two approaches (\cite{Yin2015A,Zhi2014Probe,KimXMAS,zhou2016mdash}). The rate-based algorithms first estimate the future available bandwidth according to past throughput experience, and then select the highest bitrate below the bandwidth. The buffer-based algorithms, however, only consider the current buffer conditions when making bitrate decisions. 
The hybrid methods integrate these two technologies, 
i.e., using future throughput estimates and buffer occupancy information to select the proper bitrates for future chunks. 
Based on these fixed control rules, Akhtar et al.~\cite{Akhtar:2018:OAV:3230543.3230558} dynamically adjust the configurable parameters of rules to make ABR algorithms work better over a wide range of network conditions. Xie et al.~\cite{xie2017dynamic} and Xu et al.~\cite{xu2018qoe} respectively propose a buffer-based ABR algorithm with dynamic threshold and a QoE-driven adaptive k-push algorithm for low-latency live streaming.
However, all these ABR algorithms with fixed control rules need substantial preprogramming overhead, which is not suitable for adapting to the dramatically varying channel capacity of UAVs.

Recently, there has been a growing interest in developing the optimal learning-based ABR algorithms. Before Pensieve~\cite{Mao2017Neural}, A separate line of studies~\cite{Claeys2013Design,Chiariotti2016Online,Claeys2014Design,Hooft2015A} propose to generate ABR algorithms based on reinforcement learning. However, all of these algorithms store the value function for all states instead of using value function approximation, which cannot generalize to large state and action space. Pensieve~\cite{Mao2017Neural} is the first model to apply the actor-critic network to the ABR algorithm which learns the optimal policy automatically. 
In addition, QARC~\cite{Huang:2018:QVQ:3240508.3240545} employs a DRL model to select the bitrate by jointly considering the predicted video content and network states. NAS~\cite{Yeo:2018:NAC:3291168.3291216} directly applies neural networks (NN)-based quality enhancement on the received video content to maximize the user QoE. Moreover, Jiang et al.~\cite{jiang2018chameleon} leverage the correlations of video content to dynamically pick the best configurations for analytics pipelines. Furthermore, based on the DRL algorithms, HotDASH~\cite{sengupta2018hotdash} considers the prefetching of user-preferred temporal video segments, while Guo et al.~\cite{guo2019buffer} perform dynamic resource optimization for wireless buffer-aware video streaming. Kan et al.~\cite{kan2019deep} design a DRL-based rate adaptation algorithm for 360-degree video streaming. These algorithms have their own advantages in different video transmission applications. However, they are not specifically designed for the UAV video transmission, and thus are not optimized to adapt to the dramatically changing UAV environments.


\textbf{Application of sensor data.}
The sensor data brings great convenience for IoT device communication~\cite{wang2018cross}, especially in mobile scenarios.
The sensor data is widely used in mobile devices. Several studies~\cite{Wang2016SkyEyes,Santhapuri2010Sensor, Zhang:2017:EMV:3134224.3102301,Li2014A} use sensor information to infer the motion states and surrounding environments of objects, which optimize wireless communications by adjusting protocols including client roaming, bit rate adaptation, frame aggregation and beamforming. Santhapuri et al.~\cite{Santhapuri2010Sensor} employ light sensor readings on the phone to distinguish indoor or outdoor locations and exploit the on-phone accelerometers to identify the mobility states of users, such as walking patterns and vehicular motion, which is beneficial to improve the user experience. Zhang et al.~\cite{Zhang:2017:EMV:3134224.3102301} take as input users' location information through planned routes, and then predict the bandwidth along the route to make online transmission decisions. 
Furthermore, several studies use WiFi signal strength~\cite{Krumm2004LOCADIO} or PHY layer information~\cite{Sun2015Towards} to detect the users' motion states, which are contrary to the above studies. Asadpour et al.~\cite{Asadpour2013Now} analyze in detail the impact of the relative motion between two UAVs on throughput and Wang et al.\cite{Wang2016SkyEyes} experiment to add the GPS information of the UAV to the ABR video streaming algorithm, which enables the model to indicate more UAV channel information.

\vspace{0.2cm}
\section{Conclusion}\label{sec:conclusion}

SA-ABR exploits the flight-state-related sensor data that is readily-available on today's commercial UAVs to generate ABR algorithms that provide stable and better QoE under various flight scenarios. The sensor data we analyze and exploit in experiments include GPS coordinates, accelerations, and velocities. With the help of sensor data, our model can better infer the future channel condition and effectively mitigate the negative impact caused by sudden changes in flight states. The experimental results have verified that SA-ABR outperforms the best known ABR algorithm by 21.4$\%$ in terms of average QoE reward. SA-ABR can be seamlessly integrated into existing wireless protocols and commercial hardware. We believe that with these features, SA-ABR can provide some insights for future UAV transmission policy designs.

\balance
\bibliographystyle{IEEEtran}
\bibliography{IEEEabrv,./sa-abr}

\begin{thebibliography}{10}
\providecommand{\url}[1]{#1}
\csname url@samestyle\endcsname
\providecommand{\newblock}{\relax}
\providecommand{\bibinfo}[2]{#2}
\providecommand{\BIBentrySTDinterwordspacing}{\spaceskip=0pt\relax}
\providecommand{\BIBentryALTinterwordstretchfactor}{4}
\providecommand{\BIBentryALTinterwordspacing}{\spaceskip=\fontdimen2\font plus
\BIBentryALTinterwordstretchfactor\fontdimen3\font minus
  \fontdimen4\font\relax}
\providecommand{\BIBforeignlanguage}[2]{{%
\expandafter\ifx\csname l@#1\endcsname\relax
\typeout{** WARNING: IEEEtran.bst: No hyphenation pattern has been}%
\typeout{** loaded for the language `#1'. Using the pattern for}%
\typeout{** the default language instead.}%
\else
\language=\csname l@#1\endcsname
\fi
#2}}
\providecommand{\BIBdecl}{\relax}
\BIBdecl

\bibitem{monitor}
{M}icro aerial~projects L.L.C., ``U{AV}s in agricultural; environmental
  monitoring.'' accessed July 20, 2018.

\bibitem{agri}
A.~Meola, ``Exploring agricultural drones: The future of farming is precision
  agriculture, mapping, and spraying,'' accessed Aug 1, 2017.

\bibitem{Fleureau2016POSTERGD}
J.~Fleureau, Q.~Galvane, F.-L. Tariolle, and P.~Guillotel, ``Generic drone
  control platform for autonomous capture of cinema scenes submission,'' in
  \emph{Proc. ACM MobiSys}, 2016, pp. 35--40.

\bibitem{VzrdQf}
Wired, ``Surrey now has the uk's `largest' police droneproject.'' accessed Nov
  01, 2016.

\bibitem{WwFgWk}
Techcrunch, ``Firefighting drone serves as a reminder to be careful with
  crowdfunding campaigns.'' accessed Nov 01, 2016.

\bibitem{timmurphy}
Amazon, ``Amazon {P}rime {A}ir.'' accessed Nov 01, 2016.

\bibitem{h520}
Yuneec, ``H520 overview - commercial {UAV},'' accessed 2017.

\bibitem{Stockhammer2011Dynamic}
T.~Stockhammer, ``Dynamic adaptive streaming over {HTTP} --:standards and
  design principles,'' in \emph{Proc. ACM MMSys}, 2011, pp. 133--144.

\bibitem{Jiang2012Improving}
J.~Jiang, V.~Sekar, and H.~Zhang, ``Improving fairness, efficiency, and
  stability in {HTTP}-based adaptive video streaming with festive,''
  \emph{IEEE/ACM Transactions on Networking}, vol.~22, no.~1, pp. 326--340,
  2014.

\bibitem{Sun2016CS2P}
Y.~Sun, X.~Yin, J.~Jiang, V.~Sekar, F.~Lin, N.~Wang, T.~Liu, and B.~Sinopoli,
  ``C{S}2{P}:improving video bitrate selection and adaptation with data-driven
  throughput prediction,'' in \emph{Proc. ACM SIGCOMM}, 2016, pp. 272--285.

\bibitem{Huang2014A}
T.~Y. Huang, R.~Johari, N.~Mckeown, M.~Trunnell, and M.~Watson, ``A
  buffer-based approach to rate adaptation: evidence from a large video
  streaming service,'' in \emph{Proc. ACM SIGCOMM}, 2014, pp. 187--198.

\bibitem{Spiteri2016BOLA}
K.~Spiteri, R.~Urgaonkar, and R.~K. Sitaraman, ``B{OLA}: Near-optimal bitrate
  adaptation for online videos,'' in \emph{Proc. IEEE INFOCOM}, 2016, pp. 1--9.

\bibitem{Zhi2014Probe}
Z.~Li, X.~Zhu, J.~Gahm, R.~Pan, H.~Hu, A.~C. Begen, and D.~Oran, ``Probe and
  adapt: Rate adaptation for {HTTP} video streaming at scale,'' \emph{IEEE
  Journal on Selected Areas in Communications}, vol.~32, no.~4, pp. 719--733,
  2014.

\bibitem{Yin2015A}
X.~Yin, A.~Jindal, V.~Sekar, and B.~Sinopoli, ``A control-theoretic approach
  for dynamic adaptive video streaming over {HTTP},'' in \emph{Proc. ACM
  SIGCOMM}, 2015, pp. 325--338.

\bibitem{KimXMAS}
S.~Kim and C.~Kim, ``X{MAS}: An efficient mobile adaptive streaming scheme
  based on traffic shaping,'' \emph{IEEE Transactions on Multimedia}, vol.~PP,
  pp. 1--1, 2018.

\bibitem{zhou2016mdash}
C.~Zhou, C.-W. Lin, and Z.~Guo, ``m{DASH}: A markov decision-based rate
  adaptation approach for dynamic {HTTP} streaming,'' \emph{IEEE Transactions
  on Multimedia}, vol.~18, no.~4, pp. 738--751, 2016.

\bibitem{Akhtar:2018:OAV:3230543.3230558}
Z.~Akhtar, Y.~S. Nam, R.~Govindan, S.~Rao, J.~Chen, E.~Katz-Bassett,
  B.~Ribeiro, J.~Zhan, and H.~Zhang, ``Oboe: Auto-tuning video abr algorithms
  to network conditions,'' in \emph{Proc. ACM SIGCOMM}, 2018, pp. 44--58.

\bibitem{xie2017dynamic}
L.~Xie, C.~Zhou, X.~Zhang, and Z.~Guo, ``Dynamic threshold based rate
  adaptation for {HTTP} live streaming,'' in \emph{Proc. IEEE ISCAS}, 2017, pp.
  1--4.

\bibitem{xu2018qoe}
Z.~Xu, X.~Zhang, and Z.~Guo, ``Qo{E}-driven adaptive {K}-push for {HTTP}/2 live
  streaming,'' \emph{IEEE Transactions on Circuits and Systems for Video
  Technology}, vol.~29, no.~6, pp. 1781--1794, 2018.

\bibitem{Claeys2013Design}
M.~Claeys, S.~Latre, J.~Famaey, and F.~De~Turck, ``Design and evaluation of a
  self-learning {HTTP} adaptive video streaming client,'' \emph{IEEE
  communications letters}, vol.~18, no.~4, pp. 716--719, 2015.

\bibitem{Chiariotti2016Online}
F.~Chiariotti, S.~D'Aronco, L.~Toni, and P.~Frossard, ``Online learning
  adaptation strategy for {DASH} clients,'' in \emph{Proc. ACM MMSys}, 2016,
  pp. 8:1--8:12.

\bibitem{Claeys2014Design}
M.~Claeys, S.~Latré, J.~Famaey, T.~Wu, W.~V. Leekwijck, and F.~D. Turck,
  ``Design and optimisation of a {Q}-learning-based {HTTP} adaptive streaming
  client,'' \emph{Connection Science}, vol.~26, no.~1, pp. 25--43, 2014.

\bibitem{Hooft2015A}
J.~V.~D. Hooft, S.~Petrangeli, M.~Claeys, J.~Famaey, and F.~D. Turck, ``A
  learning-based algorithm for improved bandwidth-awareness of adaptive
  streaming clients,'' in \emph{Proc. IFIP/IEEE IM}, 2015, pp. 131--138.

\bibitem{Mao2017Neural}
H.~Mao, R.~Netravali, and M.~Alizadeh, ``Neural adaptive video streaming with
  pensieve,'' in \emph{Proc. ACM SIGCOMM}, 2017, pp. 197--210.

\bibitem{Huang:2018:QVQ:3240508.3240545}
T.~Huang, R.-X. Zhang, C.~Zhou, and L.~Sun, ``Q{ARC}: Video quality aware rate
  control for real-time video streaming based on deep reinforcement learning,''
  in \emph{Proc. ACM MM}, 2018, pp. 1208--1216.

\bibitem{Yeo:2018:NAC:3291168.3291216}
H.~Yeo, Y.~Jung, J.~Kim, J.~Shin, and D.~Han, ``Neural adaptive content-aware
  internet video delivery,'' in \emph{Proc. ACM OSDI}, 2018, pp. 645--661.

\bibitem{jiang2018chameleon}
J.~Jiang, G.~Ananthanarayanan, P.~Bodik, S.~Sen, and I.~Stoica, ``Chameleon:
  scalable adaptation of video analytics,'' in \emph{Proc. ACM SIGCOMM}, 2018,
  pp. 253--266.

\bibitem{sengupta2018hotdash}
S.~Sengupta, N.~Ganguly, S.~Chakraborty, and P.~De, ``Hot{DASH}: Hotspot aware
  adaptive video streaming using deep reinforcement learning,'' in \emph{Proc.
  IEEE ICNP}, 2018, pp. 165--175.

\bibitem{guo2019buffer}
Y.~Guo, R.~Yu, J.~An, K.~Yang, Y.~He, and V.~C. Leung, ``Buffer-aware streaming
  in small scale wireless networks: A deep reinforcement learning approach,''
  \emph{IEEE Transactions on Vehicular Technology}, 2019.

\bibitem{kan2019deep}
N.~Kan, J.~Zou, K.~Tang, C.~Li, N.~Liu, and H.~Xiong, ``Deep reinforcement
  learning-based rate adaptation for adaptive 360-degree video streaming,'' in
  \emph{Proc. IEEE ICASSP}, 2019, pp. 4030--4034.

\bibitem{goodfellow2016deep}
I.~Goodfellow, Y.~Bengio, and A.~Courville, \emph{Deep learning}, 2016, vol.~1.

\bibitem{sutton2018reinforcement}
R.~S. Sutton and A.~G. Barto, \emph{Reinforcement learning: An introduction},
  2018.

\bibitem{ge2017towards}
C.~Ge, N.~Wang, G.~Foster, and M.~Wilson, ``Towards {Q}o{E}-assured 4k
  video-on-demand delivery through mobile edge virtualization with adaptive
  prefetching,'' \emph{IEEE Transactions on Multimedia}, vol.~19, no.~10, pp.
  2222--2237, 2017.

\bibitem{lu2018exploiting}
Z.~Lu, S.~Ramakrishnan, and X.~Zhu, ``Exploiting video quality information with
  lightweight network coordination for {HTTP}-based adaptive video streaming,''
  \emph{IEEE Transactions on Multimedia}, vol.~20, no.~7, pp. 1848--1863, 2018.

\bibitem{Memory2010Long}
L.~S. Memory, ``Long short-term memory,'' \emph{Neural Computation}, vol.~9,
  no.~8, pp. 1735--1780, 2010.

\bibitem{Khuwaja2018ASO}
A.~A. Khuwaja, Y.~Chen, N.~Zhao, M.-S. Alouini, and P.~Dobbins, ``A survey of
  channel modeling for uav communications,'' \emph{IEEE Communications Surveys
  \& Tutorials}, vol.~20, no.~4, pp. 2804--2821, 2018.

\bibitem{chowdhery2018aerial}
A.~Chowdhery and K.~Jamieson, ``Aerial channel prediction and user scheduling
  in mobile drone hotspots,'' \emph{IEEE/ACM Transactions on Networking},
  vol.~26, no.~6, pp. 2679--2692, 2018.

\bibitem{mnih2016asynchronous}
V.~Mnih, A.~P. Badia, M.~Mirza, A.~Graves, T.~Lillicrap, T.~Harley, D.~Silver,
  and K.~Kavukcuoglu, ``Asynchronous methods for deep reinforcement learning,''
  in \emph{Proc. ICML}, 2016, pp. 1928--1937.

\bibitem{Sutton1999Policy}
R.~S. Sutton, ``Policy gradient methods for reinforcement learning with
  function approximation,'' \emph{Advances in Neural Information Processing
  Systems}, vol.~12, pp. 1057--1063, 1999.

\bibitem{Mao2016Resource}
H.~Mao, M.~Alizadeh, I.~Menache, and S.~Kandula, ``Resource management with
  deep reinforcement learning,'' in \emph{Proc. ACM HotNets}, 2016, pp. 50--56.

\bibitem{Konda2003On}
V.~R. Konda and J.~N. Tsitsiklis, \emph{On Actor-Critic Algorithms}.\hskip 1em
  plus 0.5em minus 0.4em\relax Society for Industrial and Applied Mathematics,
  2003.

\bibitem{ChenTWLDL18}
Y.~Chen, X.~Tian, Q.~Wang, M.~Li, M.~Du, and Q.~Li, ``{ARMOR}: A secure
  combinatorial auction for heterogeneous spectrum,'' \emph{IEEE Transactions
  on Mobile Computing}, vol.~PP, pp. 1--1, 2018.

\bibitem{wang2016software}
W.~Wang, Y.~Chen, Q.~Zhang, and T.~Jiang, ``A software-defined wireless
  networking enabled spectrum management architecture,'' \emph{IEEE
  Communications Magazine}, vol.~54, no.~1, pp. 33--39, 2016.

\bibitem{wang2018cross}
W.~Wang, S.~He, L.~Sun, T.~Jiang, and Q.~Zhang, ``Cross-technology
  communications for heterogeneous iot devices through artificial doppler
  shifts,'' \emph{IEEE Transactions on Wireless Communications}, vol.~18,
  no.~2, pp. 796--806, 2018.

\bibitem{Wang2016SkyEyes}
X.~Wang, A.~Chowdhery, and M.~Chiang, ``Skyeyes: adaptive video streaming from
  {UAV}s,'' in \emph{Proc. ACM HotWireless}, 2016, pp. 2--6.

\bibitem{Santhapuri2010Sensor}
N.~Santhapuri, J.~Manweiler, S.~Sen, R.~R. Choudhury, and S.~Nelakuditiy,
  ``Sensor assisted wireless communication,'' in \emph{Proc. IEEE LANMAN},
  2010, pp. 1--5.

\bibitem{Zhang:2017:EMV:3134224.3102301}
W.~Zhang, R.~Fan, Y.~Wen, and F.~Liu, ``Energy-efficient mobile video
  streaming: A location-aware approach,'' \emph{ACM Transactions on Intelligent
  Systems and Technology}, vol.~9, pp. 6:1--6:16, 2017.

\bibitem{Li2014A}
Y.~Li, D.~Jin, Z.~Wang, P.~Hui, L.~Zeng, and S.~Chen, ``A markov jump process
  model for urban vehicular mobility: Modeling and applications,'' \emph{IEEE
  Transactions on Mobile Computing}, vol.~13.

\bibitem{Krumm2004LOCADIO}
J.~Krumm and E.~Horvitz, ``L{OCADIO}: inferring motion and location from wi-fi
  signal strengths,'' in \emph{MobiQuitous}, 2004, pp. 4--13.

\bibitem{Sun2015Towards}
L.~Sun and D.~Koutsonikolas, ``Towards motion-aware wireless lans using phy
  layer information,'' in \emph{Proc. IEEE ICNP}, 2015, pp. 467--469.

\bibitem{Asadpour2013Now}
M.~Asadpour, D.~Giustiniano, K.~A. Hummel, S.~Heimlicher, and S.~Egli, ``Now or
  later? - delaying data transfer in time-critical aerial communication,'' in
  \emph{Proc. ACM CoNEXT}, 2013, pp. 127--132.

\end{thebibliography}

\end{document}